\documentclass[12pt,preprint]{aastex}
\begin{document}
\title{Planetary nebulae in the elliptical galaxy NGC 4649 (M 60): 
 kinematics and distance redetermination\thanks{Based partly on data 
  collected at the Subaru Telescope, which is operated by the National
   Astronomical Observatory of Japan. Based partly on data collected at
    the European Southern Observatory, Chile, ESO Program 079.B-0577(A)}}

\author{A. M. Teodorescu\altaffilmark{1}, R. H. M\'endez\altaffilmark{1},
        F. Bernardi\altaffilmark{2}, J. Thomas\altaffilmark{3}, 
        P. Das\altaffilmark{3}, and O.Gerhard\altaffilmark{3}}

\email{ana@ifa.hawaii.edu, mendez@ifa.hawaii.edu}


\altaffiltext{1}{Institute for Astronomy,
      University of Hawaii, 2680 Woodlawn Drive, Honolulu, HI 96822}

\altaffiltext{2}{Universit\'a di Pisa, Largo B. Pontecorvo 5, 56127,
                 Pisa, Italy}

\altaffiltext{3}{Max Planck Institut f\"ur Extraterrestrische Physik, 
   P.O.Box 1603, D-85740 Garching bei M\"unchen, Germany}


\begin{abstract}

Using a slitless spectroscopy method with (a) the 8.2 m Subaru telescope 
and its FOCAS Cassegrain spectrograph, and (b) the ESO Very Large Telescope
(VLT) unit 1 (Antu) and its FORS2 Cassegrain spectrograph, we have 
detected 326 planetary nebulae (PNs) in the giant Virgo elliptical 
galaxy NGC 4649 (M 60), and we have measured their radial velocities. 
After rejecting some PNs more likely to belong to the companion galaxy 
NGC 4647, we have built a catalog with kinematic information for 298 PNs
in M 60. Using these radial velocities we have investigated if they support
the presence of a dark matter halo around M 60. The preliminary conclusion 
is that they do; based on an isotropic, two-component Hernquist model,
we estimate the dark matter halo mass within 3$R_{\rm e}$ to be 
4$\times10^{11} M_{\odot}$, which is almost one half of the total mass
of about $10^{12} M_{\odot}$ within 3$R_{\rm e}$. This total mass 
is similar to that estimated from globular cluster, XMM-Newton and 
Chandra observations.
The dark matter becomes dominant outside. More detailed dynamical 
modeling of the PN data is being published in a companion paper.
We have also measured the $m$(5007) magnitudes of many of these PNs,
and built a statistically complete sample of 218 PNs. 
The resulting PN luminosity function (PNLF) was used to estimate a 
distance modulus of 30.7$\pm$0.2 mag, equivalent to 14$\pm$1 Mpc. 
This confirms an earlier PNLF distance measurement, based on a much 
smaller sample. The PNLF distance modulus remains smaller than the 
surface brightness fluctuation (SBF) distance modulus by 0.4 mag. 
The reason is still unknown. 
\end{abstract}

\keywords{galaxies: distances and redshifts --- 
          galaxies: elliptical and lenticular, cD ---
          galaxies: individual (NGC 4649) ---
          galaxies: kinematics and dynamics ---  
          planetary nebulae: general ---
          techniques: radial velocities}
   
\section{INTRODUCTION}

Planetary nebulae (PNs) in the outskirts of elliptical galaxies
can be used as test particles to study dark matter existence and 
distribution (Hui et al. 1995; M\'endez et al. 2001;
Romanowsky et al. 2003; Teodorescu et al. 2005; De Lorenzi et al. 
2008, 2009; M\'endez et al. 2009; Coccato et al. 2009; Napolitano 
et al. 2010).

Sometimes the run of the line-of-sight velocity dispersion (LOSV$\sigma$)
as a function of angular distance from the center of the galaxy
shows clear evidence of dark matter; for example in NGC 5128
(Hui et al. 1995, Peng et al. 2004) or NGC 4374 (Napolitano 
et al. 2010). In other cases, the LOSV$\sigma$ shows
a Keplerian decline with distance, indicating either the absence 
of a substantial dark matter halo, or the presence of significant
radial anisotropy in the velocity distribution. Examples are
NGC 4697 (M\'endez et al. 2001, 2009; De Lorenzi et al. 2008) 
and NGC 3379 (De Lorenzi et al. 2009). 
It has been difficult to decide which interpretation
(no dark matter or radial anisotropy) is to be preferred.
The likelihood constraints from
PNs seem to weakly prefer the dark matter plus anisotropy
interpretation, but the uncertainties are large. 

One strategy to further explore this problem could be to select
an elliptical galaxy known to have a dark matter halo from other 
evidence; for example, the presence of hot, X-ray emitting gas. 
Do the PNs show a Keplerian decline of the LOSV$\sigma$ also in such a 
case? If yes, then radial anisotropy becomes the natural choice,
and we have learned about an important constraint concerning the
formation of ellipticals.

With this kind of outcome in mind, we selected the giant elliptical
M 60 (NGC 4649) in the Virgo cluster. There is abundant X-ray emitting 
gas around this E2 galaxy (Fukazawa et al. 2006; Humphrey et al. 2006;
Nagino \& Matsushita 2009; Das et al. 2010).
In addition, the globular cluster system of M 60 has been studied by 
Bridges et al. (2006), Hwang et al. (2008), and Shen \& Gebhardt (2010). 
All these studies report the presence of a dark matter halo, as expected.

A few dozens of PNs were discovered in M 60 by Jacoby et al. (1990)
as part of their effort to determine the distance to the Virgo cluster 
using the PN luminosity function (PNLF). To our knowledge their radial 
velocities were never measured.

In this paper we present the results of a search for PNs in several 
fields around M 60. Section 2 describes the observations, reduction 
procedures, PN detection and photometry. Section 3 deals with the 
slitless radial velocity method, its calibration and results. 
In Section 4 we present a catalog with all detected sources, and
discuss which ones belong to the companion galaxy, the spiral NGC 4647.
In section 5 we discuss rotation and the run of LOSV$\sigma$ as a function 
of angular distance from the center of M 60, make a first attempt to fit 
it using Hernquist models, and analyze a plot of the escape 
velocity as a function of angular distance. A more detailed dynamical
analysis of M 60, based on the N-body made-to-measure code NMAGIC 
(De Lorenzi et al. 2007) is being published separately 
(Das et al. 2011). Section 6 is devoted to a redetermination of the 
distance to M 60 using the PNLF, and section 7 is a recapitulation.

\section{OBSERVATIONS, REDUCTIONS, DETECTIONS, AND PHOTOMETRY}

Planetary nebulae can be detected in the light of [O III] $\lambda$5007 
using the traditional on-band, off-band filter technique. Having taken 
the on-band image, insertion of a grism as dispersing element produces 
not only dispersion but also a displacement of all images. 
The dispersed images of PNs remain point sources, 
which permits an accurate measurement of the displacement. Calibration 
of the displacement as a function of wavelength and position in the CCD 
offers an efficient way of measuring radial velocities for all 
PNs in the field, irrespective of their number and distribution. We have 
described the method in several previous papers (M\'endez et al. 2001,
2009; Teodorescu et al. 2005, 2010).

Because of the rather large angular size of M 60, with an effective 
radius $R_{\rm e} = 128$ arcsec (Kormendy et al. 2009), we decided to 
observe several fields around M 60. Figure 1 shows the fields we  
observed, using the ESO Very Large Telescope (VLT) at the Cerro Paranal, 
Chile, and the Subaru telescope at Mauna Kea, Hawaii, USA.
Chronologically, we first observed Field 1 in April 2007 with Subaru; 
because of poor seeing we reobserved Field 1, and added Field 2, in June
2007, with the VLT; finally, we reobserved Field 1 and added Field 3, in
May 2008, with Subaru.

\subsection{VLT Observations}      

Images and grism slitless spectra of M 60 were taken with the FORS2 
spectrograph attached to the Cassegrain focus of unit telescope UT1 
(Antu) of the ESO VLT, on the first halves of the nights of 2007 
June 9/10 and 10/11. These half-nights were reasonably 
clear, but not of photometric quality, with thin cirrus clouds passing
sometimes. The average seeing during these 2007 nights was 0.7 arcsec.
FORS2, with the standard-resolution collimator, gave a field
of 6.8$\times$6.8 arc minutes on a mosaic of two 2k$\times$4k MIT CCD's
(pixel size 15 $\mu$m). All images were binned 2$\times$2.
The image scale was 0.125 arcsec pixel$^{-1}$ before binning. The 
on-band and off-band
interference filters had the following characteristics:
effective central wavelengths, in observing conditions,
of 5028 and 5300 \AA ; peak transmissions of 0.76 and
0.80; equivalent widths of 48.5 and 215 \AA ; and FWHMs of 60 and 250 \AA.
The dispersed images were obtained with the holographic grism 1400V. This 
grism gave a dispersion of 21 \AA \ mm$^{-1}$, or 0.31 \AA \ pixel$^{-1}$, 
at 5000 \AA, before binning, and obviously 0.62 \AA \ pixel$^{-1}$ after
binning.

Table 1 lists the most important FORS2 science images obtained for this 
project, with the corresponding exposure times and air masses.

\subsection{Subaru Observations}

Images and grism slitless spectra of M 60 were taken with the Faint Object
Camera and Spectrograph (FOCAS) attached to the Cassegrain focus of the 
8.2 m Subaru telescope, Mauna Kea, Hawaii, on the nights of 2007 April 
16 and 17, and of 2008 May 3 and 4. All these nights were of 
photometric quality, with the only exception of 2007 April 17, affected by 
some thin cirrus. The average seeing during the 2007 nights was
one arcsec, and 0.6 arcsec in 2008.

FOCAS has been described by Kashikawa et al. (2002). The field of view of 
FOCAS is 6.5 arc minutes and was covered by two CCDs of 2k  
$\times$ 4k (pixel size 15  $\mu$m) with an unexposed gap of 5$''$ 
between them. All images were binned 2$\times$1 in the horizontal direction
(perpendicular to dispersion) to increase the signal from the very faint 
sources we want to detect, but without compromising the spectral 
resolution.
The image scale is 0.104 arcsec pixel$^{-1}$ after unbinning.
The on-band filter has a central wavelength of 5025 \AA, a FWHM of 60 \AA, 
a peak transmission of 0.68 and an equivalent width of 40 \AA.
Off-band imaging was done through the standard broadband visual filter.
The dispersed images were obtained inserting an echelle grism with 175 
grooves mm$^{-1}$ which operates in the 4th order and gives a 
dispersion of 0.5 \AA \ pixel$^{-1}$, with an efficiency of 60\%.

The spectrophotometric standard G 138-31 (Oke 1990) was used for the 
photometric calibration of the on-band images.
Table 2 lists the most important FOCAS science images obtained for this 
project, with the corresponding exposure times and air masses.

\subsection{Data reductions}

Standard IRAF\footnote{IRAF is
distributed by the National Optical Astronomical Observatories,
operated by the Association of Universities for Research in
Astronomy, Inc., under contract to the National Science
Foundation} tasks were used for the basic CCD reductions 
(bias subtraction, and flat field correction using twilight flats).
In order to eliminate the cosmic-rays and to detect faint PN candidates, 
we needed to combine the scientific images of M 60 for each different
field. The procedure we followed for this image combination has been 
carefully described in previous papers (e.g. M\'endez et al. 2009, 
Teodorescu et al. 2010). For brevity, we do not repeat the description 
here.

For easier PN detection and photometry in the central parts of M60, where 
the background varies strongly across the field, we produced difference 
images between undispersed on-band and off-band combined frames. In ideal 
conditions, this image subtraction should produce a flat noise frame with 
the emission-line sources as the only visible features. 
A critical requirement to achieve the desired result is 
perfect matching of the point-spread functions
(PSFs) of the two frames to be subtracted. For this purpose, we applied a 
method for ``optimal image subtraction'' developed by Alard \& Lupton 
(1998), and implemented in Munich by G\"ossl \& Riffeser (2002) as part 
of their image reduction pipeline.

This procedure cannot be used for the combined dispersed images, because 
there is no off-band counterpart. Therefore, to flatten the background and 
reduce the contamination by stellar spectra, we applied the 
IRAF task ``fmedian'' to the combined dispersed images.
The resulting median images were then subtracted from the unmedianed ones. 

\subsection{PN detection and slitless spectroscopy}

The PNs we want to find have strong [O III] $\lambda$5007 emission and 
extremely weak, essentially absent continuum. 
We identified the PNs by blinking the on-band versus the off-band 
combined images, and confirmed them by blinking on-band versus dispersed. 
In addition, the object had to be a point source, and had to be 
undetectable in the off-band image. In this way it is possible to 
minimize the contamination of the PN sample by unrelated background 
sources, like galaxies with emission lines redshifted into the on-band
filter transmission curve. We will further discuss 
contamination issues in Section 4. The ($x$ and $y$) pixel 
coordinates of all the PN candidates in the undispersed 
and dispersed images were measured with the IRAF task ``phot'' with the 
centering algorithm ``centroid''.

We performed an astrometric calibration of the images using the
USNO-B1 astrometric star catalog (Monet et al. 2003), using a method we 
have described in Teodorescu et al. (2010). 
We estimate an rms of about 0.3 arcsec for the astrometry of our PNs.

\subsection{Photometry}

The purpose of our photometry was to obtain magnitudes $m$(5007)
as defined by Jacoby (1989):

\begin{equation}
m(5007)=-2.5 \ {\rm log} I(5007) - 13.74. \label{mdef}
\end{equation}

where $I$(5007) is the [O III] 5007\AA \ flux measured through 
the on-band filter. 
Since the FORS2 data were taken under non-photometric conditions, 
we restricted our photometric measurements to the FOCAS data, 
which were enough for our purposes (a redetermination of the PNLF 
distance). 
For the flux calibration, we adopted the
spectrophotometric standard star G 138-31 (Oke 1990). 
This star has a monochromatic flux at 5025 \AA \ of 
1.51 $\times$ 10$^{-15}$ 
erg cm$^{-2}$ s$^{-1}$ \AA$^{-1}$ (Colina \& Bohlin 1994). 
The flux measured through the on-band filter, in units of 
ergs cm$^{-2}$ s$^{-1}$, can be calculated knowing the 
equivalent width of the on-band filter; using equation (1), we find 
$m$(5007)=19.31 for G 138-31.

Many PNs were measurable only on the differences of the combined images 
(on $-$ off). Thus, to calculate the $m$(5007) of the PNs we had to go through 
several steps. First, we made aperture photometry of G 138-31 using the IRAF 
task ``phot''. We also made aperture photometry of four moderately bright 
stars in the reference images selected for image registering.
These four ``internal standards'' were selected 
relatively distant from the center of M 60, to avoid background problems. 

Having tied the spectrophotometric standard to the internal frame standards, 
we switched to strictly differential photometry. We made aperture 
photometry of the internal standards on the on-band 
combined images to correct for any differences relative to the reference 
images. On the same on-band 
combined images we subsequentely made PSF-fitting DAOPHOT photometry 
(Stetson 1987; IRAF tasks ``phot'', ``psf'' and ``allstar'') of the internal 
standards and four bright PNs. From the aperture photometry and PSF-fitting 
photometry of the internal standards we determined the aperture correction. 
Finally, we made PSF-fitting photometry of all PN candidates on the difference 
images (onband $-$ offband), where the stars had been eliminated. The four 
bright 
PNs were used to tie this photometry to that of the standards. The internal 
errors in the photometry of the difference images were estimated to be below 
5\%. Applying a final correction related to the peak of the on-band filter 
transmission curve (see Jacoby et al. 1987), we obtained $m$(5007) for 
272 PNs.

\section{RADIAL VELOCITY CALIBRATIONS AND RESULTS}

We have introduced changes in the calibration procedure required for the
FORS2 data. Our original method (M\'endez et al. 2001) used FORS1, where
multiple slits could be defined using pairs of hanging blades. FORS2 allows
to use especially designed masks covering the focal surface, making it
possible to switch to a more convenient calibration method, which is the 
one that we used with FOCAS (M\'endez et al. 2009). For brevity, we will
limit our description to the FOCAS calibration.

In order to determine the shift produced by the insertion of the grism
as a function of wavelength and position on the CCD, we used a special
mask that produces an array of point sources when it is illuminated with
the internal calibration lamps or any extended astronomical source. 
The full mask
is made up of almost 1000 calibration points, separated by almost 100 pixels.
We also used exposures of the local PN Abell 35 (PNG 303.6+40.0). 
This PN has a large 
angular size that allowed us to obtain calibration measurements all across 
the FOCAS field. An example of these calibration images is shown in Figure 2 
of M\'endez et al. (2009). The procedure for wavelength measurement is 
explained in Section 3 of that paper, to which we refer the interested 
reader.

For the wavelength calibration, after each grism exposure, the special mask
was inserted in the light path, and on-band and grism + on-band images 
were obtained, illuminating the mask with the comparison lamp. 
In addition, on-band and grism + on-band images of the special mask
were taken illuminating the mask with Abell 35, for radial velocity
quality control. 

To test for the presence of any systematic errors in the radial 
velocities, we use the images of Abell 35 as follows. We can 
measure radial velocities in two ways: 
(a) classical, using each mask hole as a slit, and (b) slitless, 
using the displacement introduced by the grism
as a measure of wavelength, and therefore velocity.
The comparison between slitless and classical measurement is shown
in Figures 2 and 3 for Chip 1 and Chip 2, respectively. Since we are 
measuring the velocity of different gas elements in Abell 35, we expect 
to get a range of velocities across the field. 
There is good agreement between classical and slitless measurements,
indicating no problems with the calibration of the slitless method.
The average heliocentric velocity of Abell 35 from all the grid points 
is about $-$12 km s$^{-1}$, in good agreement with 
measurements of the systemic velocity ($-16 \pm 14$ km s$^{-1}$, 
Jacoby 1981; $-7 \pm 4$ km s$^{-1}$, Schneider et al. 1983).
We conservatively estimate the calibration errors in FOCAS 
slitless radial velocities to be of the order of 10 km s$^{-1}$. 
If we add quadratically the uncertainties in velocity given by the 
calibration errors ($\sim$ 10 km s$^{-1}$), the position errors 
($\sim$ 10 km s$^{-1}$), and the errors from image registration 
($\sim$ 10 km s$^{-1}$), we get an overall error of about 
17 km s$^{-1}$. 
Assuming that the spectrograph deformations and guiding errors have 
a marginal contribution, we estimate the total uncertainty in the 
velocities measured with Subaru and FOCAS to be at most 20 km s$^{-1}$.
The expected errors in FORS2 velocities are a bit larger, around 25
km s$^{-1}$, because of the slightly lower spectral resolution after 
binning. In Figure 4 we show a comparison of velocities for 50 objects 
that were measured with both spectrographs. We find a standard deviation 
of 40 km s$^{-1}$, which is not unreasonable, given the error bars 
reported for both instruments. Having found sufficiently good 
agreement, all available measurements for each object were averaged 
when building the catalog. 

\section{CATALOG OF PNs IN M 60}

We have found 326 PN candidates in our search. Figure 5 shows them all.
The $x$, $y$ coordinates are measured in arcsec and are relative to the 
center of M 60. The $x$ coordinate runs in the direction of increasing RA
along the major axis of M 60, defined to be at P.A.=105$^\circ$ (from N 
through E). The $y$ coordinate runs in the direction of increasing 
Declination.

There is a complication introduced by the presence of the companion 
spiral galaxy NGC 4647. Some of our detections must belong to NGC 4647. 
Figure 6a is a histogram showing number of PNs as a function of velocity.
There is a clear peak at the velocity of NGC 4697, which is 1409 
km s$^{-1}$, according to the NASA/IPAC Extragalactic Database (NED).
The peak is easily noticed mostly because the dispersion in radial 
velocity of this face-on spiral is very low.

In order to eliminate this source of contamination, we used a 
combination of two criteria (see McNeil et al. 2010). The first 
criterion is the relative surface brightness 
contribution from the two galaxies at the position of each PN. The
surface brightness profile of M 60 was taken from Kuchinski et al. 
(2000) and Lee et al. (2008); that of NGC 4647 is from Kuchinski et al.
The fit to M 60 was a $r^{1/4}$ profile, and the disk of NGC 4647 was 
fitted with an exponential law. The ratio of fluxes at the position of 
the PN was assumed to give the ratio of probability of membership 
based on light alone. The second and most important criterion is 
the relative likelihood that an object with a given velocity belongs 
to the velocity distribution of M 60 versus that of NGC 4647. The 
velocity distributions were approximated by Gaussians,
and the probability of membership was calculated using the algorithm 
of Ashman et al. (1994). Since the NGC 4647 Gaussian is so narrow 
(100 km s$^{-1}$), this is the dominant factor. We have decided to 
reject 28 PNs that are less than 20 times more likely to belong to 
M 60 than to NGC 4647, according to the two combined criteria we have 
just described. Figure 6b shows the histogram for the remaining 298 
objects; the NGC 4647 peak has disappeared.

Table 3 lists the 298 PNs we have accepted as belonging to M 60. 
Table 4 lists the 28 remaining PNs, which more probably (but not 
necessarily) belong to NGC 4647. Figure 7 shows the distribution of 
PNs near NGC 4647, indicating which ones were deleted from the M 60 
list.

It may be useful to explain the naming 
strategy in Tables 3 and 4. Objects identified with an ID number $<$300 
appear in only one FOCAS field. If 600 $<$ ID $<$ 999, the object has 
been detected in two FOCAS fields. If 1000 $<$ ID $<$ 1999, the object 
is present in only one FORS2 field. If 2000 $<$ ID $<$ 2999, the object
is present in FORS2 Field1 Chip2, and Field2 Chip1. 
Objects seen with both FOCAS and FORS2 keep their 
FOCAS ID, unless no FOCAS velocity could be measured. In other words,
objects having only a FORS2 velocity are identified with the FORS2 ID.

Figures 8 and 9 show the 298 M 60 PN velocities as functions of the 
$x$-coordinate in arcsec and the $y$-coordinate in arcsec, respectively.
The average velocity of 1065 km s$^{-1}$ is in reasonable agreement, within 
the uncertainties, with the NED radial velocity of 1117 km s$^{-1}$ for 
M 60. We estimate the PN average velocity uncertainty to be approximately 
30 km s$^{-1}$, from a velocity dispersion of the order of 300 km s$^{-1}$
(see next section), the number of PNs measured, and the possible systematic 
error of $\pm$ 10 km s$^{-1}$ in our velocities from the calibration 
procedure using Abell 35. The 28 PNs we have assigned to NGC 4647 give an 
average velocity of 1447 km s$^{-1}$, again in reasonable agreement with 
the corresponding NED velocity of 1409 km s$^{-1}$ for NGC 4647. But of 
course this was to be expected, because velocity was one of the two 
arguments we used to decide whether or not any given object could belong 
to NGC 4647. 

In addition to PN sample contamination from NGC 4647, which we discussed
above, we can also consider possible contamination by background
galaxies having some emission line that has been redshifted into the
on-band filter transmission curve. Given the spectral resolution of 
0.5 \AA \ per pixel, the radial velocity resolution is about 140 
km s$^{-1}$; i.e., the PN internal velocity field is not resolved, 
and PNs appear as point sources. An emission line from a contaminating 
galaxy might be broader, but not necessarily so. Of course 
such a broad line would be immediately rejected as a PN,
because it would not be a point source; therefore,
we do not have that kind of contamination. But not all
contaminating emission-line galaxies have necessarily
broad lines; and so there could be contamination by
narrow-emission galaxies. Knowing the central wavelength 
(5025 \AA) and the FWHM (60 \AA), it is easy to
calculate, using the Doppler formula, the radial velocity 
range for detection of [O III] $\lambda$5007 (from $-$600 km s$^{-1}$ 
to 2700 km s$^{-1}$). The observed radial velocity range is much 
narrower (from 200 to 1700 km s$^{-1}$, see Figures 8 and 9). 
Therefore, first, we are not missing any PNs; and second, contamination
by background galaxies is not important, because such contaminants would 
be expected to be randomly distributed in wavelength across the filter 
transmission curve. Instead, we only find sources at the expected 
redshift of M 60, with the expected dispersion, and no discrepant 
velocities at all. 

\section{PRELIMINARY ANALYSIS OF M 60 RADIAL VELOCITIES}

A deeper study of the kinematic information is deferred to another
paper (Das et al. 2011). Here we limit ourselves to presenting a 
preliminary analysis, based on simple methods we have used previously 
to analyze other galaxies.
Our main purpose is to study the run of the line-of-sight velocity 
dispersion (LOSV$\sigma$). Since our measurement of the LOSV$\sigma$ 
at large angular 
radii can be affected by rotation effects, we start by quantifying how
important the rotation effects can be.

\subsection{Rotation}

We start with rotation along the major axis. Restricting our sample to 
those PNs with $|y| < 100$ arcsec, we defined 5 groups with increasing 
average $x$ coordinate, and calculated the average velocity for each 
group. The result is shown in Figure 10. We find that, on average, 
negative $x$ objects are receding and positive $x$ approaching, which is 
in agreement with the observed behavior of absorption-line data (Fisher 
et al. 1995, Pinkney et al. 2003) and globular clusters (Hwang et al.
2008). 
There is also some evidence of rotation in the perpendicular direction,
i.e. along the minor axis. In Figure 9 we see two groups of PNs,
with $y < -200$ and $y > 200$, respectively, with average velocities
higher (1138 km s$^{-1}$; 11 objects) and lower (974 km s$^{-1}$; 25 
objects) than the systemic velocity of M 60. This behavior is not 
observed in the globular cluster sample of Hwang et al. (2008), so
its significance is not clear. We defer a discussion of rotation 
to the accompanying paper of Das et al. (2011). However, the existence
of a velocity gradient as a function of the $y$ coordinate will be 
taken into account when we discuss the behavior of the LOSV$\sigma$.

\subsection{Line-of-sight velocity dispersion}

In order to study the LOSV$\sigma$, we subdivided our 298 PN sample into a 
central region and three elliptical annuli, with shapes similar to that 
of M 60, at increasing angular distances from the center of M 60.
The numbers of PNs per region, from the inside out, are 75, 75, 74,
and 74, respectively. The annuli are shown in Figure 11.
For all PNs within each region we calculated the average 
angular distance to the center, and the LOSV$\sigma$. 
The result of these calculations is shown in Figure 12. The LOSV$\sigma$
derived from the PNs (four data points) is compared with 
that derived from major axis, long-slit, absorption-line spectra 
of M 60 (Fisher et al. 1995, Pinkney et al. 2003). We find 
reasonable agreement, within error bars, between PNs and 
absorption-line data within 60$''$ of the galaxy's center. We have 
added three more PN data points, obtained by isolating PNs with large 
positive $y$ coordinates (25 objects), large negative $y$ coordinates 
(11 objects), and large positive $x$ coordinates (37 objects). The idea
was to test if the differential motion of these groups around the center 
of M 60 can affect the measured LOSV$\sigma$. But the effect does not seem 
to be significant, with the only exception of the group with large 
positive $x$ coordinates, which indeed gives a lower LOSV$\sigma$. 
Note that we do not have a similar number of PNs on the other side 
of the galaxy (large negative $x$ coordinates) because of the 
presence of NGC 4647.

Let us consider what all this information tells us about dark matter 
existence and distribution in M 60. We start by making a rough 
estimate of the visible mass in this galaxy, from its blue luminosity 
and the $(M/L)_B$ ratio, which can be estimated from evolutionary 
population synthesis models. This has been done in the recent 
literature. If we follow, for example,
Shen \& Gebhardt (2010), using the tables and figures of
Maraston (1998, 2005), we obtain an estimated $(M/L)_B$ = 10
(this is based on a Salpeter initial mass function).
Adopting a distance modulus of 31.1 mag (Blakeslee et al. 2009),
an extinction-corrected $B_T$ = 9.7, and the solar $B$ absolute 
magnitude 5.47, we obtain for M 60 a blue luminosity of
6 $\times 10^{10}$ $L_{\odot}$, which then gives a mass of 
6 $\times 10^{11}$ $M_{\odot}$.

We now ask if this visible mass is enough to explain the large 
central LOSV$\sigma$ in Figure 12. To investigate this, we use an analytical
model by Hernquist (1990). This model is spherical, non-rotating, 
isotropic, and it assumes a constant mass-to-light ratio.
We adopt an effective radius $R_{\rm e}$ = 128$''$ (Kormendy 
2009), which is equivalent to 10.5 kpc for a distance of 17 Mpc,
and a total mass of 6 $\times 10^{11}$ $M_{\odot}$. 
The fit (dotted line in Figure 12) obviously fails. 
In order to fit the central LOSV$\sigma$ we need a mass of 
1.15 $\times 10^{12}$ $M_{\odot}$ (dashed line). But
even this larger mass, assuming always a constant $M/L$ ratio,
cannot explain the large LOSV$\sigma$ at angular distances larger than 150 
arcsec. A better fit can be obtained using a two-component Hernquist
mass distribution, as in Hui et al. (1995):

\begin{equation}
M(r) = \frac{M_l \ r^2}{(r+a)^2} + \frac{M_d \ r^2}{(r+d)^2}
\end{equation}

\noindent where $M_l$ and $M_d$ are the visible and dark matter total
masses, and $a$ and $d$ are the corresponding scale lengths. Given the
corresponding density and potential, we compute the projected
velocity dispersion as a function of distance from the center 
using a code that numerically integrates the Jeans equation, assuming
isotropic orbits, and expands the resulting three-dimensional and 
projected dispersions in Chebyshev polynomials.
The two-component Hernquist model (solid line in Figure 12) 
successfully fits the observed LOSV$\sigma$, if we adopt the 
following parameters: 
$M_l = 9.3\times10^{11} \ M_{\odot}$,
$M_d = 6.6\times10^{12} \ M_{\odot}$,
and $d = 17 \ a$ (note that in the Hernquist model, 
$R_{\rm e} = 1.8153 \ a$). We do not attribute a lot of significance
to the numerical values of these parameters; we are content with the
implication that there seems to be a dark matter halo in the outskirts
of M 60, as expected. However, for easier comparison with other studies 
in the literature, we will give a few specific numbers. From Eq. (2)
we find that the dark matter halo mass within 3$R_{\rm e}$ is
4$\times10^{11} \ M_{\odot}$, which is almost one half of the 
total mass of about $10^{12} \ M_{\odot}$ within 3$R_{\rm e}$. 
This total mass within 3$R_{\rm e}$ is similar to that estimated 
from XMM-Newton and Chandra observations; see, e.g., Figure 8 in 
Nagino \& Matsushita (2009). The dark matter becomes dominant outside.
The total mass within 3$R_{\rm e}$ derived by Shen \& Gebhardt (2010) 
from globular cluster studies is a bit higher, namely a few 
times $10^{12} \ M_{\odot}$ (see their Figure 4).

Of course our use of Hernquist models is just a 
first approximation; in another paper (Das et al. 2011) we present
a more careful dynamical analysis of M 60, based on the N-body 
made-to-measure code NMAGIC (De Lorenzi et al. 2007). Our purpose
in the present paper is to make the PN database available, and make a 
first exploration of its significance.

In Figure 12 we find evidence suggesting that a significant fraction
of the matter within a few $R_{\rm e}$ must be dark. This is in
interesting contrast to the evidence in other galaxies like NGC 4697
and NGC 821 (M\'endez et al. 2009, Teodorescu et al. 2010), where the
one-component, constant mass-to-light ratio Hernquist model did provide
a good fit to the run of the LOSV$\sigma$ for a normal $(M/L)_B$ of 10 or 
less. Leaving that aside, the PN kinematics have passed the M 60 test; 
pending a more detailed study, we conclude in principle that, where 
independent evidence indicates the presence of an extended dark matter
halo, the PNs give compatible results.

\subsection{Escape velocities}

We can make a complementary test, by plotting PN radial velocities as 
a function of
angular distance from the center of M 60. Suppose we compare with the 
local escape velocity for the lower-mass Hernquist model with constant 
mass-to-light ratio used in Figure 12. The escape velocity is given by: 

\begin{equation}
V_{\rm esc} = (2 G M_{\rm t} / (r + a))^{0.5}, 
\end{equation}

\noindent where $M_{\rm t}$ is the total (stellar, visible) mass, and 
$a$ is the scale length equal to $R_{\rm e}/1.8153$. In the presence 
of a substantial dark matter contribution, we would expect some PNs 
to show velocities in excess of the escape velocity calculated from 
the visible mass, as shown in NGC 5128 by Hui et al. (1995) and Peng 
et al. (2004). Figure 13 shows that this is indeed the case in M 60. 
When we increase the mass to the level required to explain the 
LOSV$\sigma$, we no longer find unbound PNs.

\section{THE PNLF, DISTANCE, AND PN FORMATION RATE}

There is an unexplained discrepancy between PNLF distances 
and distances derived from the method of surface brightness 
fluctuations (SBF; Tonry et al. 2001). This problem has been described
by Ciardullo et al. (2002), and a recent update is provided by 
Teodorescu et al. (2010). Since the original PNLF distance to M 60
(Jacoby et al. 1990) was based on a small sample of 17 PNs, in fact 
the smallest sample used in their paper, we decided it was worthwhile
to make a PNLF distance redetermination using our much larger sample.

Having measured the apparent magnitudes $m$(5007) of 272 PNs, 
we needed to produce a 
statistically complete sample, because the detectability of a PN varies 
with the background brightness. For this purpose we used a procedure that
has been described in Section 5 of M\'endez et al. (2001). In summary,
we eliminated all PNs fainter than $m$(5007) = 27.6, beyond which 
the number of PNs per bin shows a marked decrease, indicating severe 
incompleteness; and we also eliminated all PNs within a zone of exclusion
characterized by high background counts and more difficult detectability.
This zone of exclusion was an ellipse at the center of M 60, with minor
and major semiaxes of 40 and 50 arcsec, respectively. In this way, we got 
a statistically complete sample of 218 PNs.

The PN luminosity function (PNLF) was built, using 0.2 mag bins, 
and compared with simulated PNLFs like those used by 
M\'endez \& Soffner (1997) to fit the observed PNLF of M 31. The comparison
is shown in Figure 14. The absolute magnitudes $M$(5007) that produce
the best fit to the simulated PNLF were calculated using 
an extinction correction of 0.09 mag at 5007 \AA \ (from data listed 
in the NED; see Schlegel et al. 1998) and a distance modulus 
$m - M$ = 30.7, which is equivalent to a distance of 14 Mpc. 
The simulated PNLFs plotted in Figure 14 are binned, like the observed one, 
into 0.2 mag intervals, and have maximum final mass of 0.63 $M_{\odot}$, 
$\mu_{\rm max}$ = 1, and sample sizes between 2400 and 
6700 PNs (see M\'endez \& Soffner 1997; the ``sample size'' is the total 
number of PNs, detected or not, that exist in the surveyed area). Since 
the observed PNLF shows an evident change of slope in Figure 14, 
the fit gives unambiguous information about both the distance 
modulus and the sample size.
We estimate an error of 0.1 mag from the goodness of the fit at 
different distance moduli. To obtain the total error estimate, we have to 
combine the possible systematic and random errors. The 
systematic error is the same as in Jacoby et al. (1990), i.e., 0.13 mag, 
including the possible error in the distance to M 31, in the modeling of the 
PNLF and in the foreground extinction. The random contributions are given by 
0.1 mag from the fit to the PNLF, 0.05 mag from the photometric zero point, 
and 0.05 mag from the filter calibration.
Combining all these errors quadratically, we estimate that the total error 
bar for the PNLF distance modulus is $\pm$0.2 mag. The PNLF distance 
modulus, 30.7, is smaller than the SBF distance modulus (31.1, according
to Tonry et al. 2001, and Blakeslee et al. 2009).

Our new PNLF distance is in excellent agreement with the previous PNLF 
distance estimate (Jacoby et al. 1990). Obviously, to obtain a reliable 
PNLF distance we do not need to observe so many PNs as we have done in 
this work. However, observing large samples opens the possibility of 
studying the {\it shape} of the PNLF in more detail than previously 
possible. Since we still lack a thorough understanding of the reason
why the bright end of the PNLF is so insensitive to stellar population 
differences (e.g. Ciardullo 2006, M\'endez et al. 2008), more empirical
information about PNLF shapes could offer vital clues for progress. 

One of the suggested ways of solving the PNLF-SBF discrepancy is by
contamination of the PNLF with background galaxies that have some 
emission line redshifted into the on-band filter transmission curve. 
Teodorescu et al. (2010) have shown that this idea does not work; in 
several cases (NGC 1344, NGC 821, NGC 4697) we have the necessary 
amount of kinematic information to rule out a large number of 
contaminants. Since we have not found higher-redshift contaminants 
in M 60, we now have one more argument against the PNLF contamination 
idea.

Once we know the sample size (about 4000, from Figure 14), we can 
calculate the specific PN formation rate $\dot{\xi}$ in units of 
PNs yr$^{-1}$ $L_{\odot}$$^{-1}$, 

\begin{equation}
n_{\rm PN} = \dot{\xi} L_{\rm T} t_{\rm PN},
\end{equation}

where $n_{\rm PN}$ is the sample size, $L_{\rm T}$ 
is the total bolometric luminosity of the sampled population, 
expressed in $L_{\odot}$, and $t_{\rm PN}$ is the 
lifetime of a PN, for which 30,000 yr was adopted 
in the PNLF simulations. 
Adopting an extinction-corrected $B_T$ = 9.7, $B-V$ = 0.95, a
bolometric correction of $-$0.85 (Buzzoni et al. 2006), and a solar
absolute bolometric magnitude 4.74, we obtain for M 60 a luminosity
of 1.5$\times 10^{11} \ L_{\odot}$, of which we have sampled some 
8$\times 10^{10} \ L_{\odot}$, if we discount the elliptical 
zone of exclusion at the center, and the area immediately 
near NGC 4647. Therefore, with $n_{\rm PN} = 4000$, we obtain 
$\dot{\xi} = 1.7\times 10^{-12}$ PNs yr$^{-1}$ $L_{\odot}$$^{-1}$.
We can also express the PN formation rate as 
$\alpha = n_{\rm PN} / L_{\rm T}$. Using that definition, we find
log $\alpha$ = $-$7.3, in good agreement with the previous determination,
as can be seen most easily in Figures 11 and 12 of Buzzoni et al. 
(2006).

The fact that we have found PNs associated with NGC 4647, with similar 
Jacoby magnitudes, tells us that its PNLF distance will be similar to 
that of M 60, as expected. We could have tried to obtain a more
quantitative determination of the distance to NGC 4647 from the PNs 
in Table 4, but we have refrained from doing so, because the result 
would be unreliable. First of all, the 
number of objects is rather small. In addition, we might suffer
extra uncertainty from extinction related to NGC 4647, which we 
cannot estimate. Finally, the NGC 4647 PNLF may be contaminated 
with a few M 60 PNs; some of the PNs that we have listed in 
Table 4 as ``more likely to belong to NGC 4647" can conceivably 
belong instead to M 60, which is the dominant galaxy in this 
pair.

\section{SUMMARY OF CONCLUSIONS}

We have discovered many new PNs in M 60, and have built a catalog of 
298 PNs with measured radial velocities. 
A preliminary study of the LOSV$\sigma$ derived 
from this database indicates that some dark matter must be present
within one $R_{\rm e}$, and that there must be an extended dark matter 
halo, as expected from previous independent evidence. Thus, the
kinematic information provided by the PNs proves to be qualitatively 
consistent, in
principle, with inferences based on X-ray and globular cluster data.
A more detailed dynamical study, using the N-body made-to-measure code 
NMAGIC (De Lorenzi et al. 2007), lies beyond the scope of this paper, 
and is being published separately (Das et al. 2011).

Photometry of most of the PNs has been used to build a statistically 
complete sample of 218 PNs. The resulting luminosity function has been
used for a redetermination of the PNLF distance, finding excellent 
agreement with the earlier PNLF distance determination of Jacoby et al. 
(1990), which was based on a much smaller sample. The new PNLF distance 
modulus remains 0.4 mag smaller than the SBF distance modulus (Blakeslee 
et al. 2009).

This work was supported by the National Science Foundation (USA) under 
grants 0307489 and 0807522. In our research we made use of the NASA/IPAC
Extragalactic Database (NED), which is operated by the Jet Propulsion 
Laboratory, California Institute of Technology, under contract with the 
National Aeronautics and Space Administration. It is a pleasure to 
acknowledge the help provided by the Subaru staff, in particular the 
support astronomer Takashi Hattori. Similar thanks are extended to the 
VLT staff at Cerro Paranal. We thank Jan Kleyna for making available a 
computer program required to compute the projected velocity dispersion 
as a function of distance from the center for a two-component Hernquist
mass distribution. We thank the anonymous referee for some useful 
suggestions.

\clearpage

\figcaption[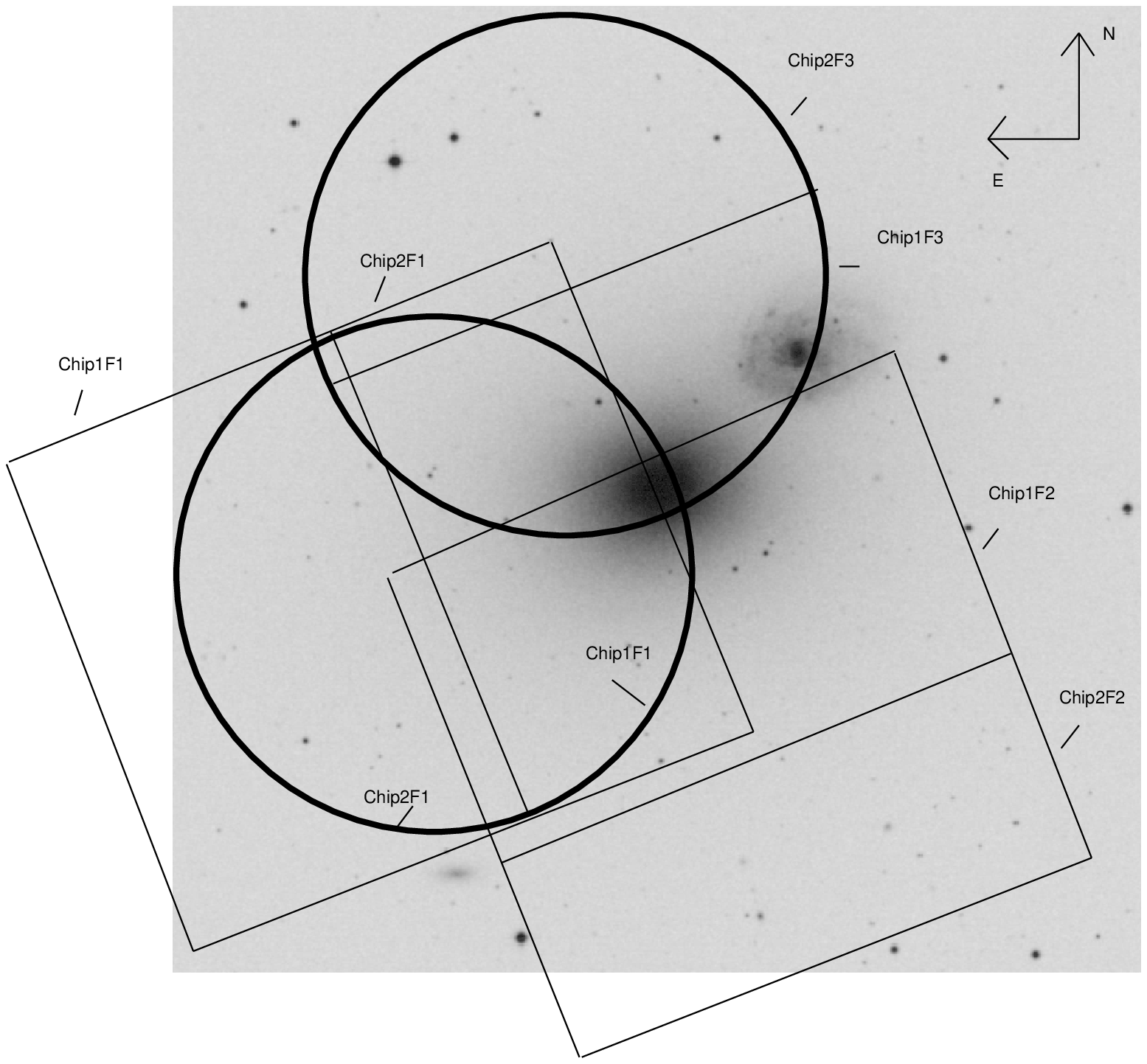]{Fields F1, F2, F3 observed for this project. The 
circular fields correspond to FOCAS, and the rectangular ones to FORS2. 
The spiral galaxy is NGC 4647. The total area of the sky shown here is 
12.9$\times$12.9 arc minutes.
\label{fig1}}

\figcaption[f02.ps]{Comparison of slitless vs slit (classical) radial 
velocities across Abell 35 for Chip 1 of FOCAS.
\label{fig2}}

\figcaption[f03.ps]{Comparison of slitless vs slit (classical) radial 
velocities across Abell 35 for Chip 2 of FOCAS.
\label{fig3}}

\figcaption[f04.ps]{Comparison of heliocentric slitless velocities 
measured with both FORS2 and FOCAS for 50 PN candidates. 
We plot FORS2 velocities (left) and (FOCAS $-$ FORS2) 
velocities (right) as a function of FOCAS velocities.
We obtain good agreement, with a standard deviation of 40 
km s$^{-1}$.
\label{fig4}}

\figcaption[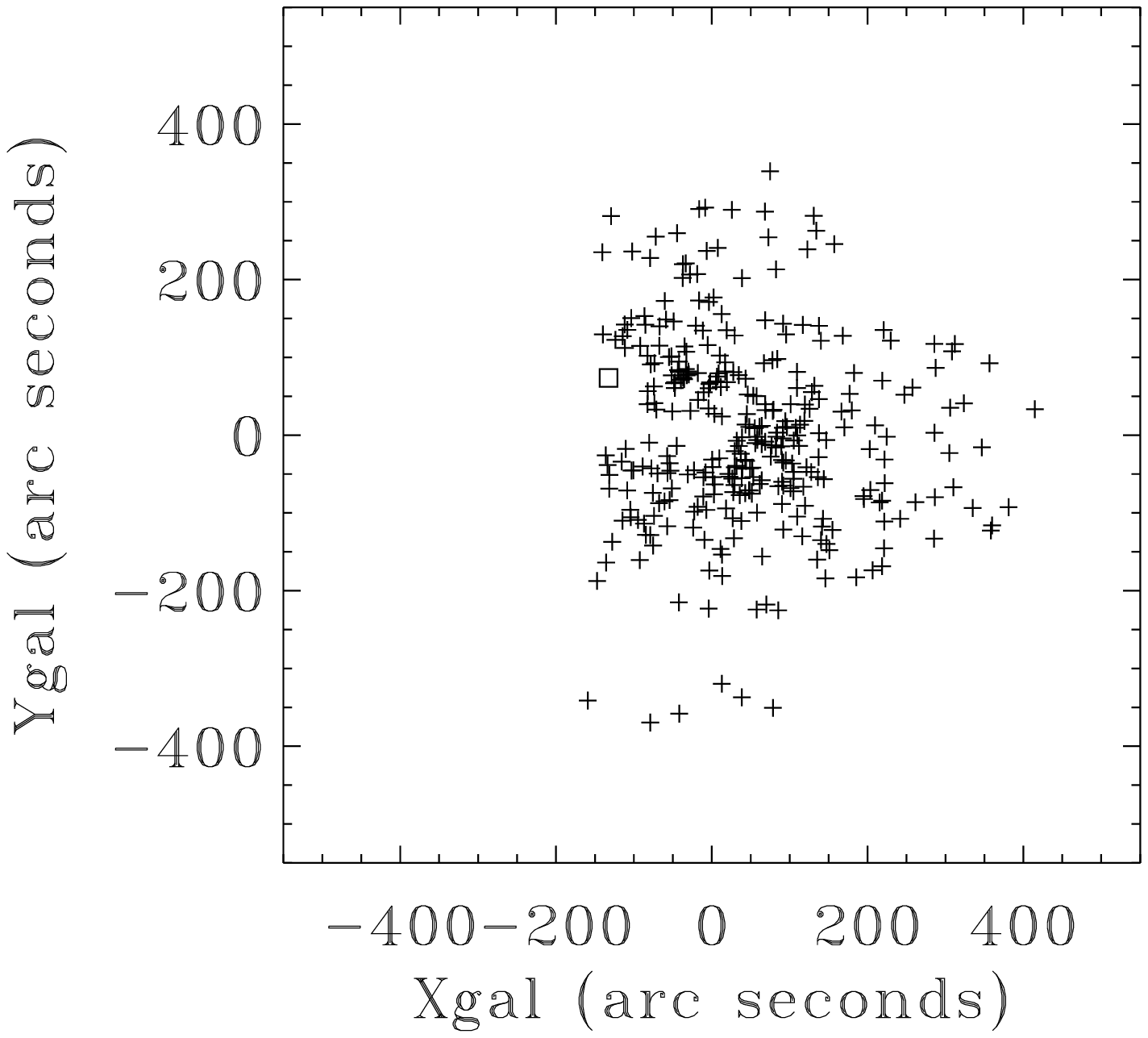]{Positions $x, y$ in arc seconds, relative to the 
center of M 60, for the 326 PN candidates with measured 
velocities (plus signs). The square marks the position of 
the center of NGC 4647. The $x$ coordinate runs in the direction 
of increasing RA, along the major axis of M 60. The $y$ 
coordinate runs along the minor axis, in the direction of increasing 
Declination. Because of these choices, the sky appears flipped from 
left to right.
\label{fig5}}

\figcaption[f06.ps]{{\it Left}: Number of PNs as a function of velocity 
for 326 detections. There is a peak at the velocity of NGC 4647
(1409 km s$^{-1}$), indicating some contamination. 
{\it Right}: Number of PNs as a function of velocity for the 
298 detections assigned to M 60 (see the text). 
The NGC 4647 peak has disappeared.
\label{fig6}}

\figcaption[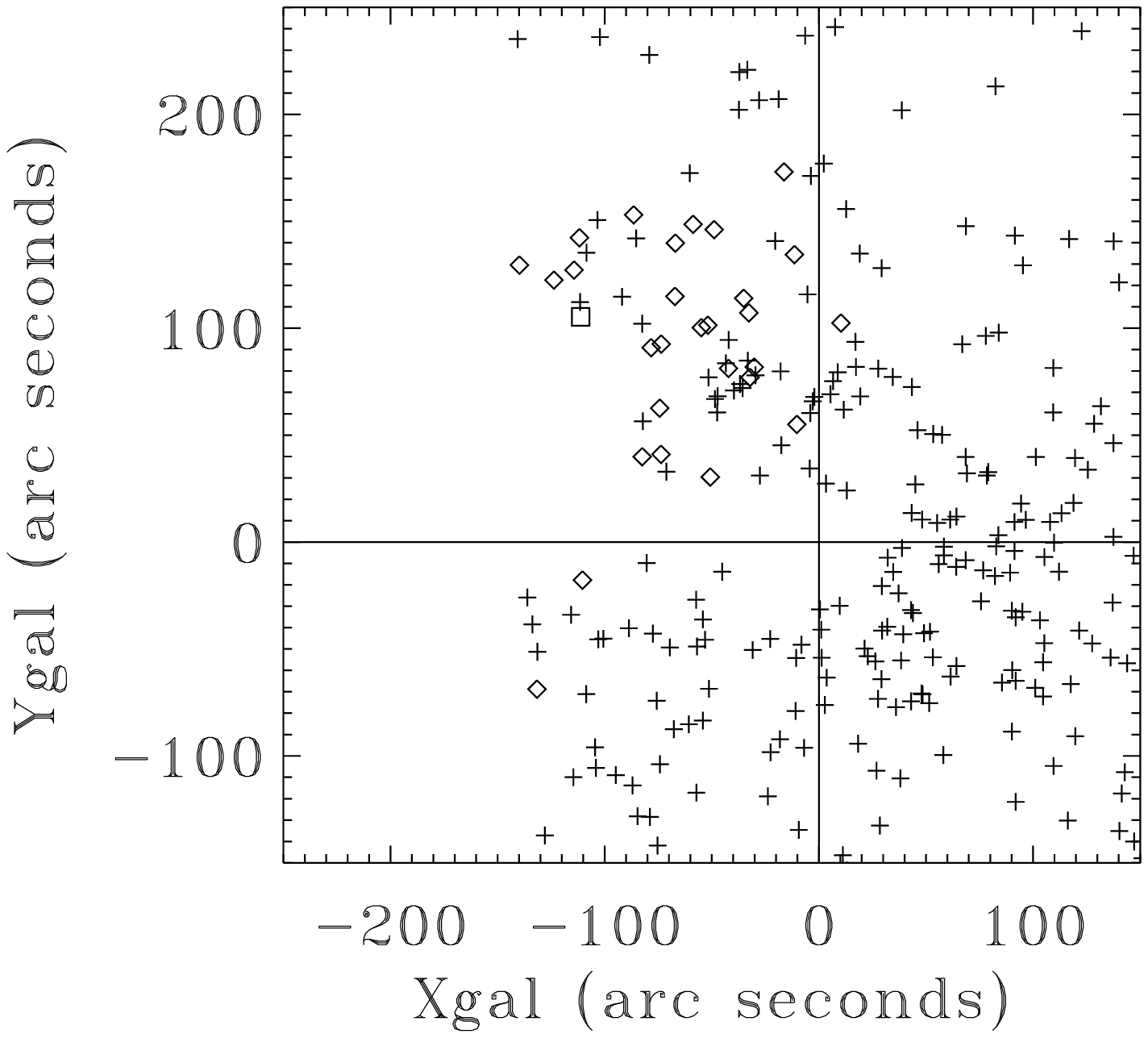]{Positions $x, y$ in arc seconds, relative to the 
center of M 60, for PNs near NGC 4647. The square
indicates the center of NGC 4647. Objects rejected from the total sample, 
because they may belong to NGC 4647, are indicated with diamonds. Plus
signs represent the M 60 PN sample.
\label{fig7}}

\figcaption[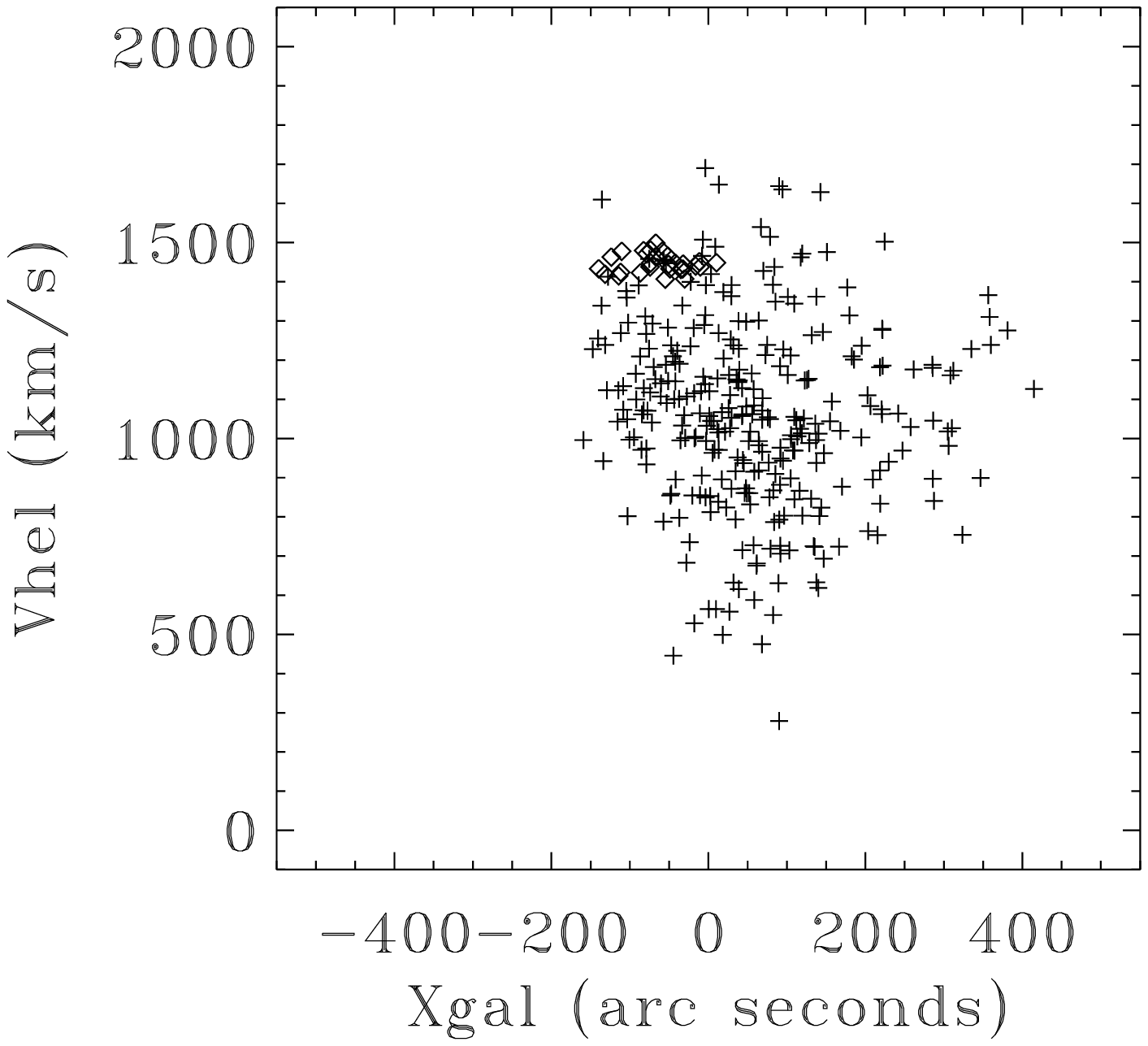]{Plus signs: heliocentric radial velocities of the 
298 PNs as a function of their $x$-coordinates in arc seconds 
relative to the center of M 60. Diamonds: 28 objects listed in Table 4.
\label{fig8}}

\figcaption[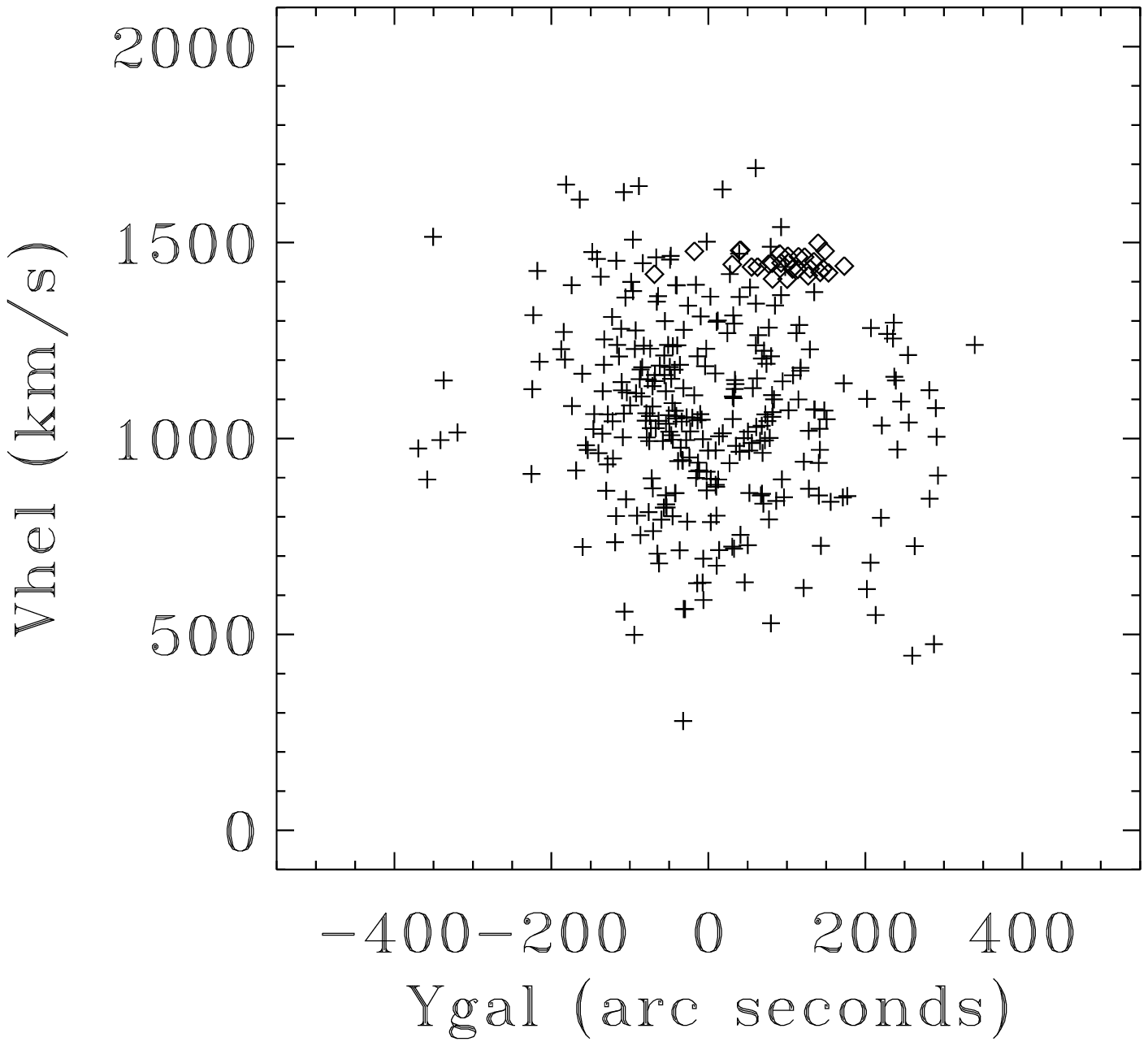]{Plus signs: heliocentric radial velocities of the 
298 PNs as a function of their $y$-coordinates in arc seconds 
relative to the center of M 60. Diamonds: 28 objects listed in Table 4.
\label{fig9}}

\figcaption[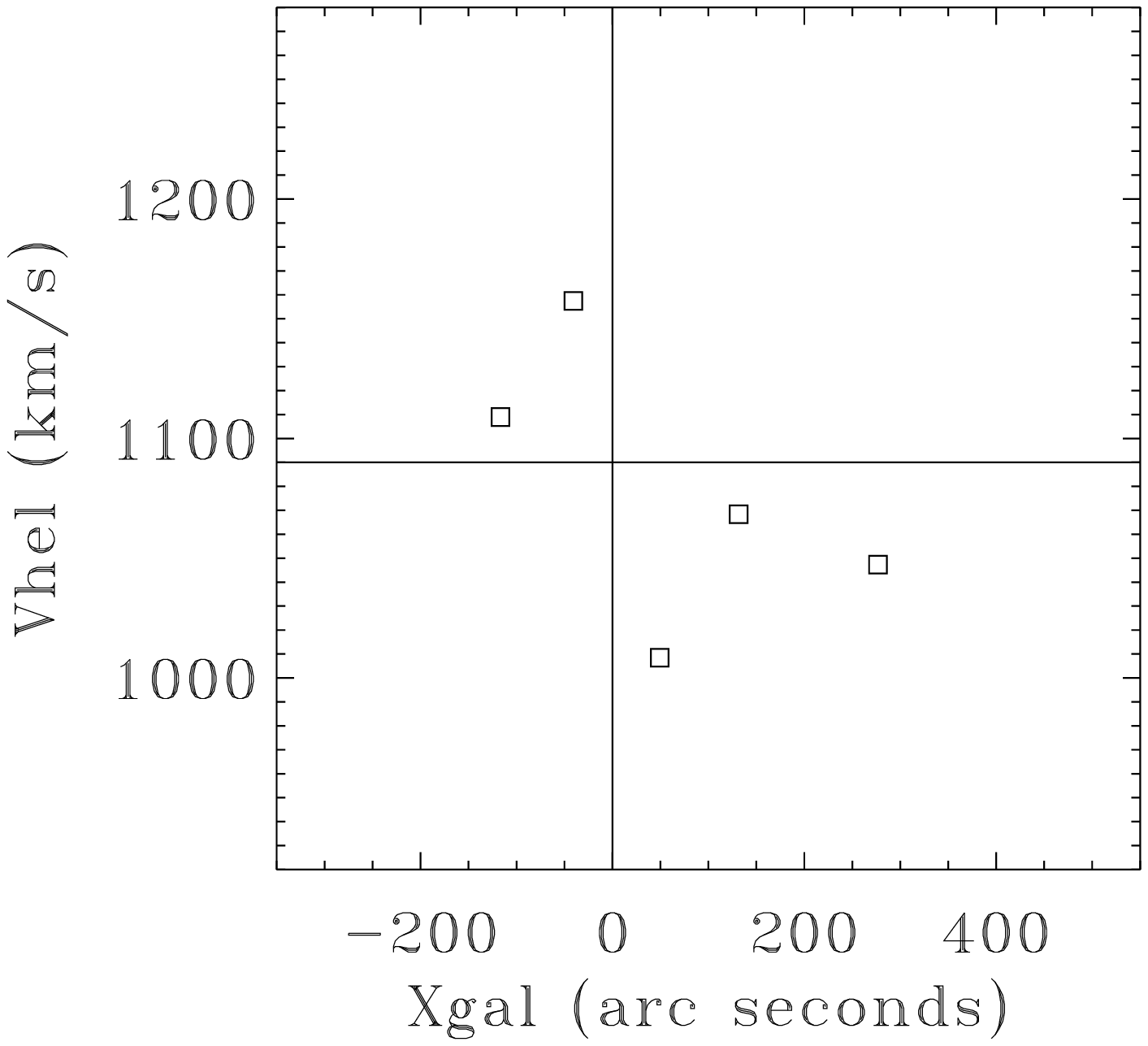]{Average velocities plotted as a function of average 
$x$ coordinate for five PN groups, defined in the text. The numbers of 
objects in each group, from left to right, are 8, 42, 84, 35 and 24. 
\label{fig10}}

\figcaption[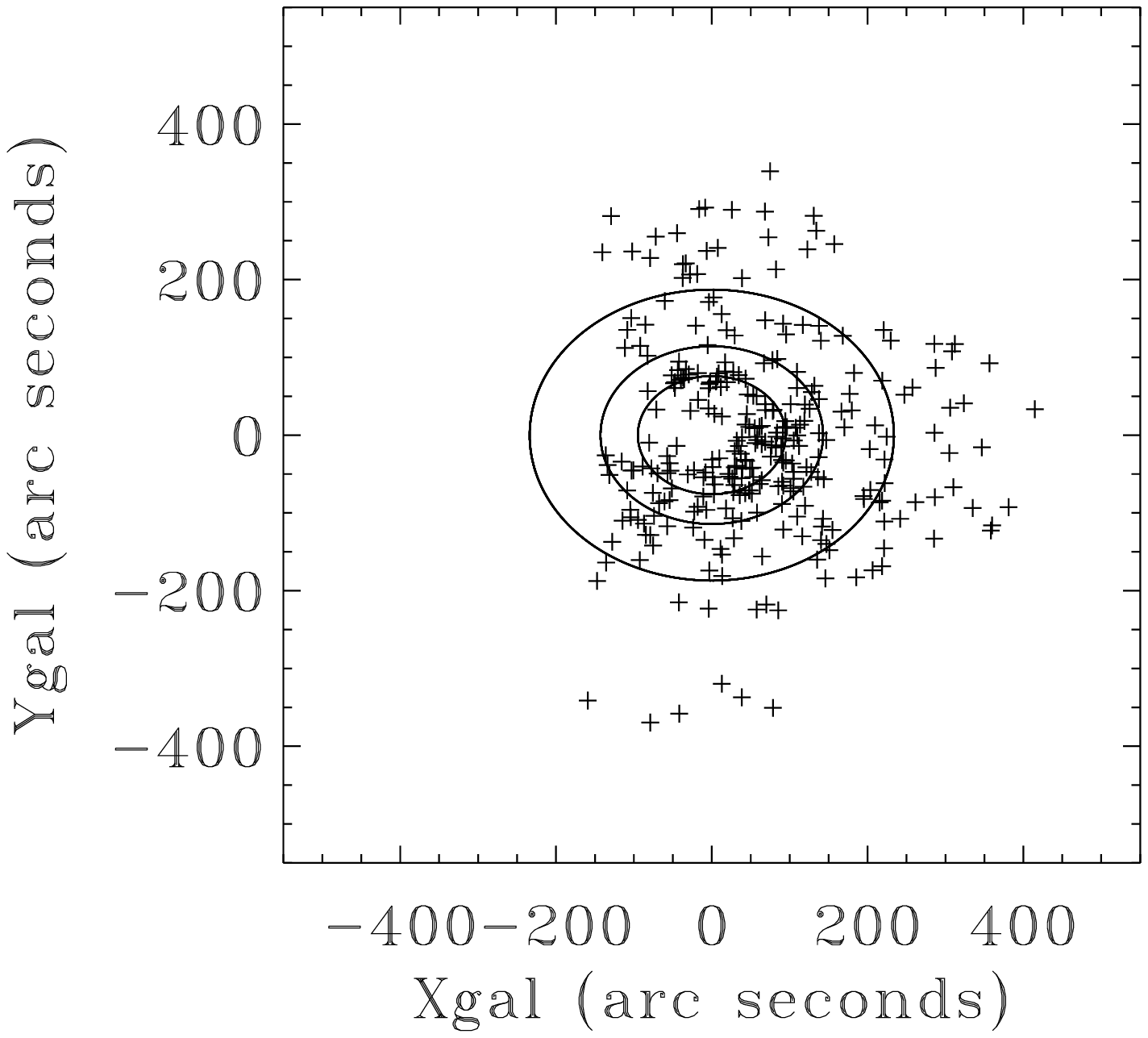]{The elliptical annuli used to divide the plane of 
the sky into four regions, for the calculation of LOSV$\sigma$s. Plus 
signs represent the 298 PNs in the M 60 sample.
\label{fig11}}

\figcaption[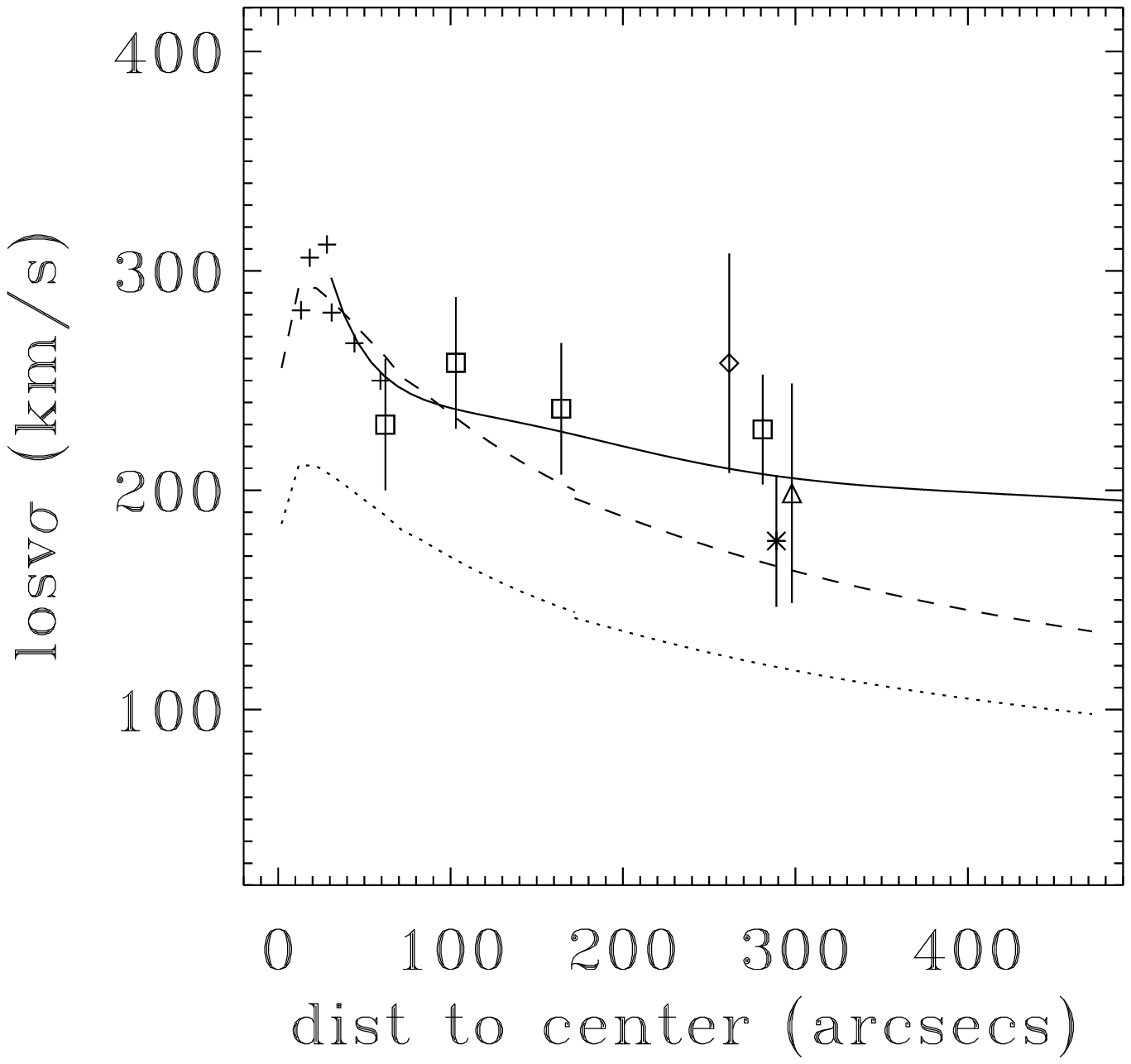]{LOSV$\sigma$ plotted as a function of 
average angular distance to the center of M 60.
The PNs were divided into 4 regions, as explained in the text.
These 4 data points are represented as squares. 
In addition, there is a diamond for PNs with large positive $y$,
a triangle for PNs with large negative $y$, and an asterisk for PNs
with large positive $x$. Plus signs are major axis, long-slit 
absorption-line data (Fisher et al. 1995, Pinkney et al. 2003). 
The dotted line represents the analytical model of 
Hernquist (1990), with a constant $M/L$ ratio, a total mass of 
6 $\times$ 10$^{11}$ $M_{\odot}$, and $R_{\rm e}$= 128$''$.
The dashed line is the same kind of model, but with a higher mass of
1.15 $\times$ 10$^{12}$ $M_{\odot}$. The solid line is a two-component
Hernquist mass distribution, as described in the text.
\label{fig12}}

\figcaption[fig13.ps]{Individual PN radial velocities plotted as a 
function of angular distance from the center of M 60. The solid 
lines are escape velocities for Hernquist models with total masses
1.2 x 10$^{12}$ $M_{\odot}$ (outer lines) and 6 x 10$^{11}$ $M_{\odot}$ 
(inner lines).
\label{fig13}}

\figcaption[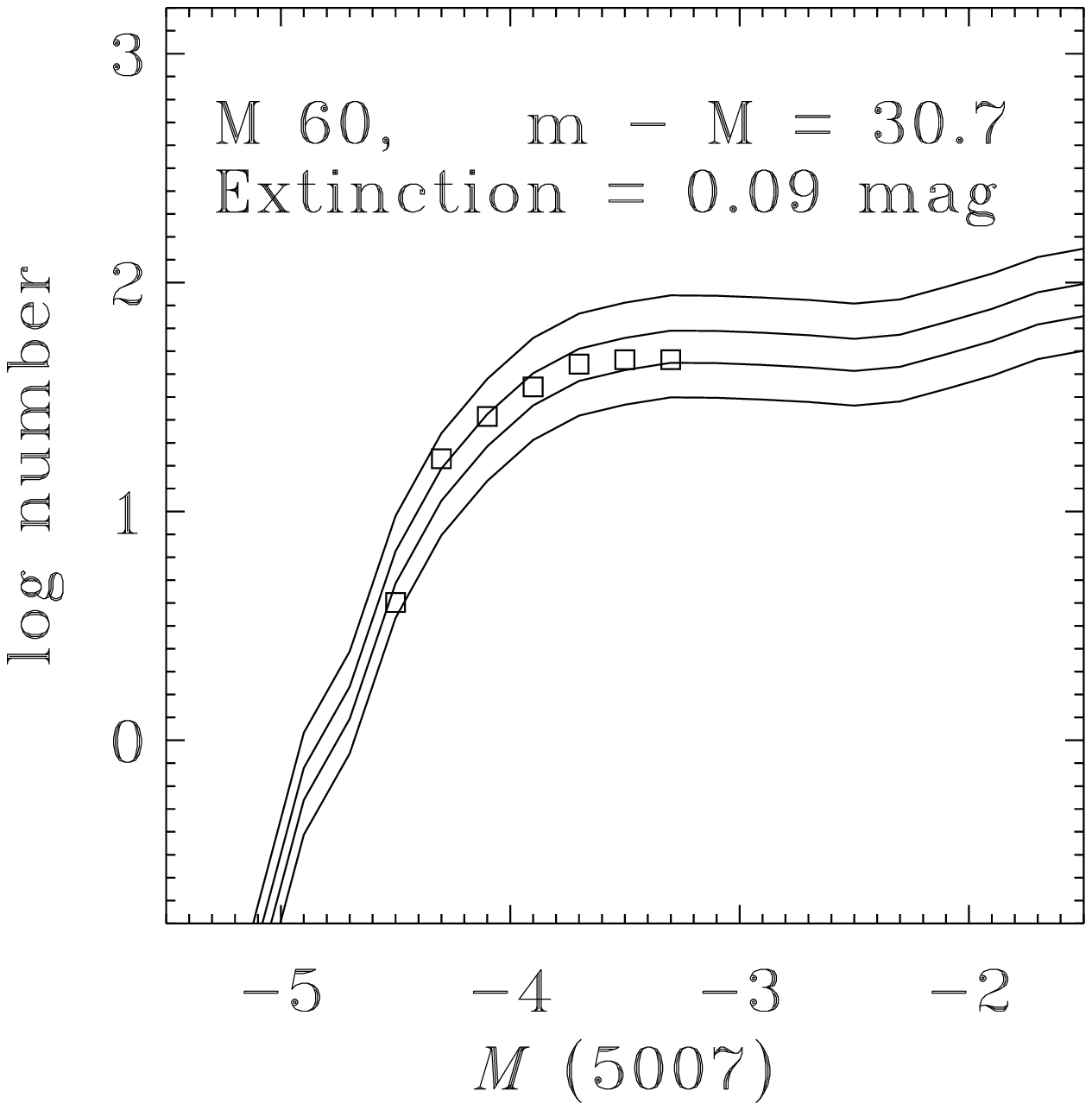]{Observed [O III] $\lambda$5007 PNLF of M 60 
(squares), with the statistically complete sample of 218 PNs binned 
into 0.2 mag intervals. The apparent magnitudes m(5007) have been 
transformed into absolute magnitudes M(5007) by adopting an extinction 
correction of 0.09 mag and a distance modulus $m - M$ = 30.7. 
The four lines are PNLF simulations (M\'endez and Soffner 1997) for 
four different total PN population sizes: 2400, 3400, 4700, and 6700 PNs. 
We estimate the best-fit sample size to be 4000. From this sample size 
it is possible to estimate the PN formation rate (see the text).
\label{fig14}}


\clearpage

\begin{figure}
\figurenum{1}
\epsscale{1.0}
\plotone{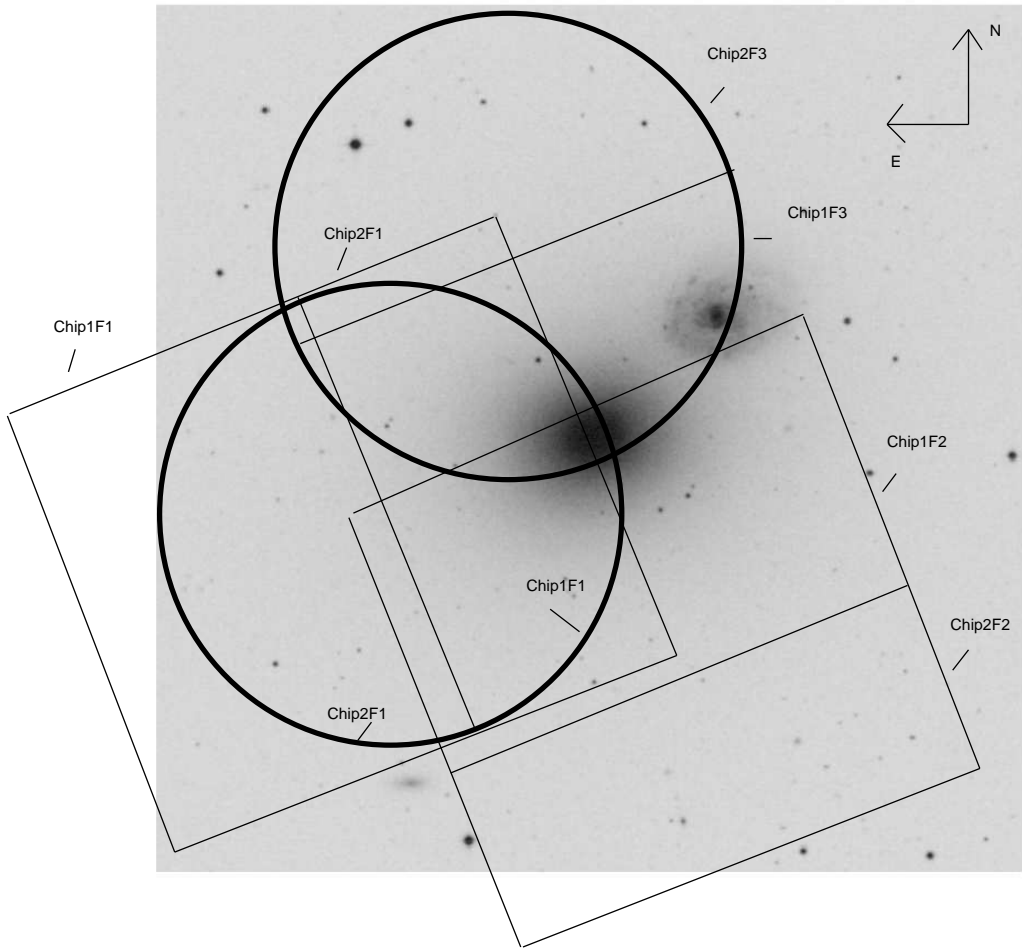}
\caption{Fields F1, F2, F3 observed for this project. The circular fields 
correspond to FOCAS, and the rectangular ones to FORS2. The spiral 
galaxy is NGC 4647. The total area of the sky shown here is 
12.9$\times$12.9 arc minutes.
}
\end{figure}

\begin{figure}
\figurenum{2}
\epsscale{1.0}
\plottwo{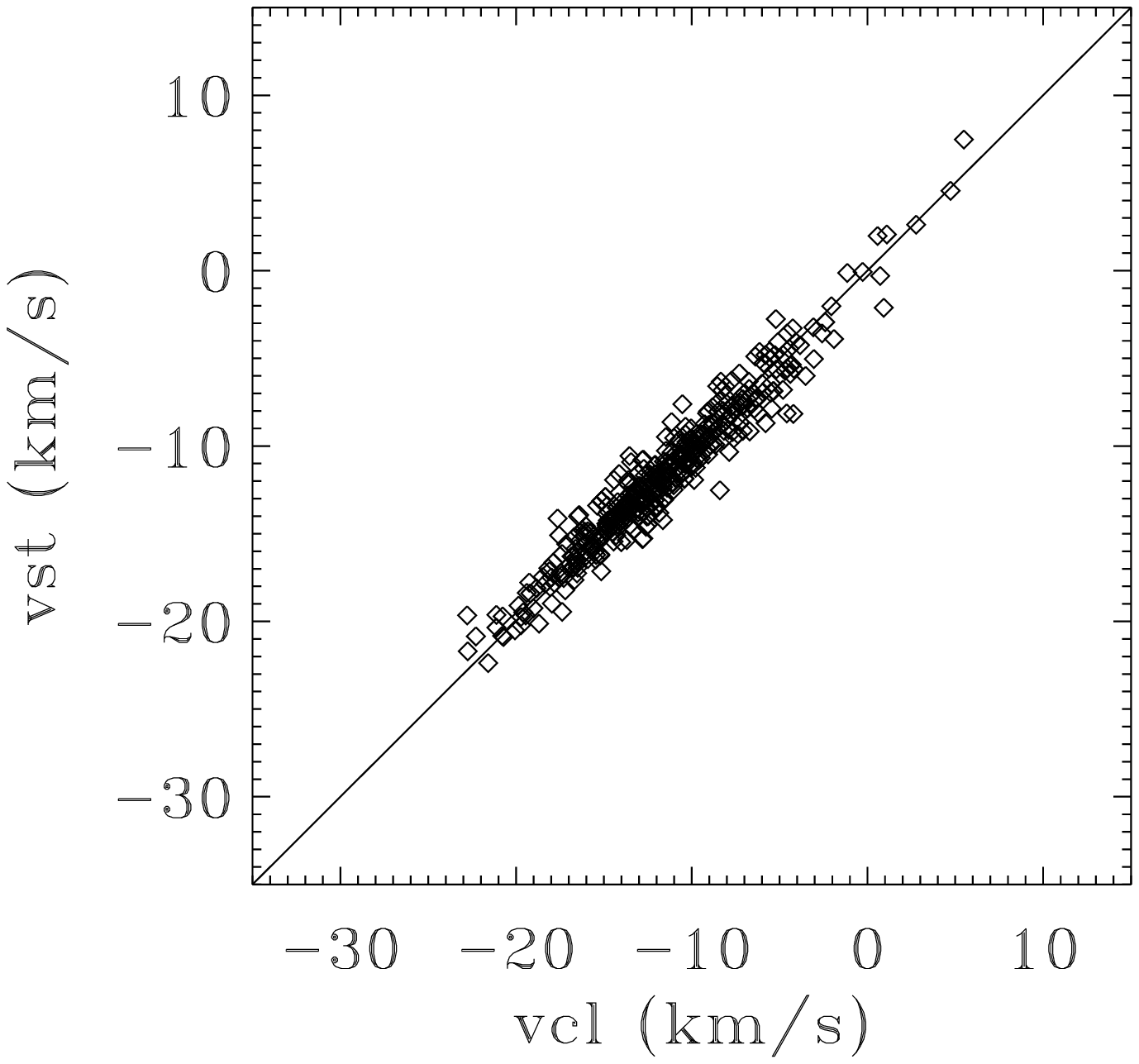}{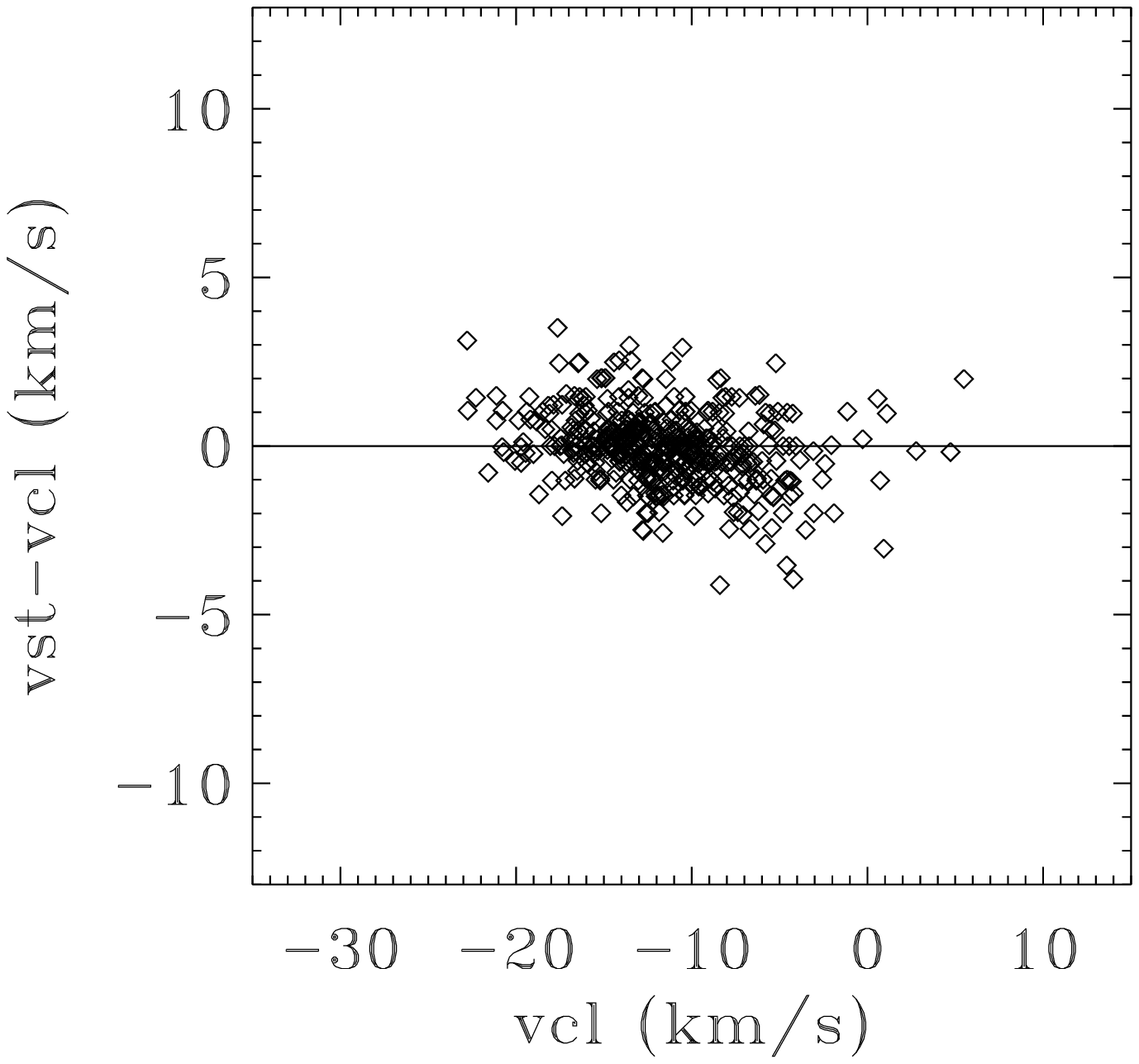}
\caption{Comparison of slitless vs slit (classical) radial 
velocities across Abell 35 for Chip 1 of FOCAS.
}
\end{figure}

\begin{figure}
\figurenum{3}
\epsscale{1.0}
\plottwo{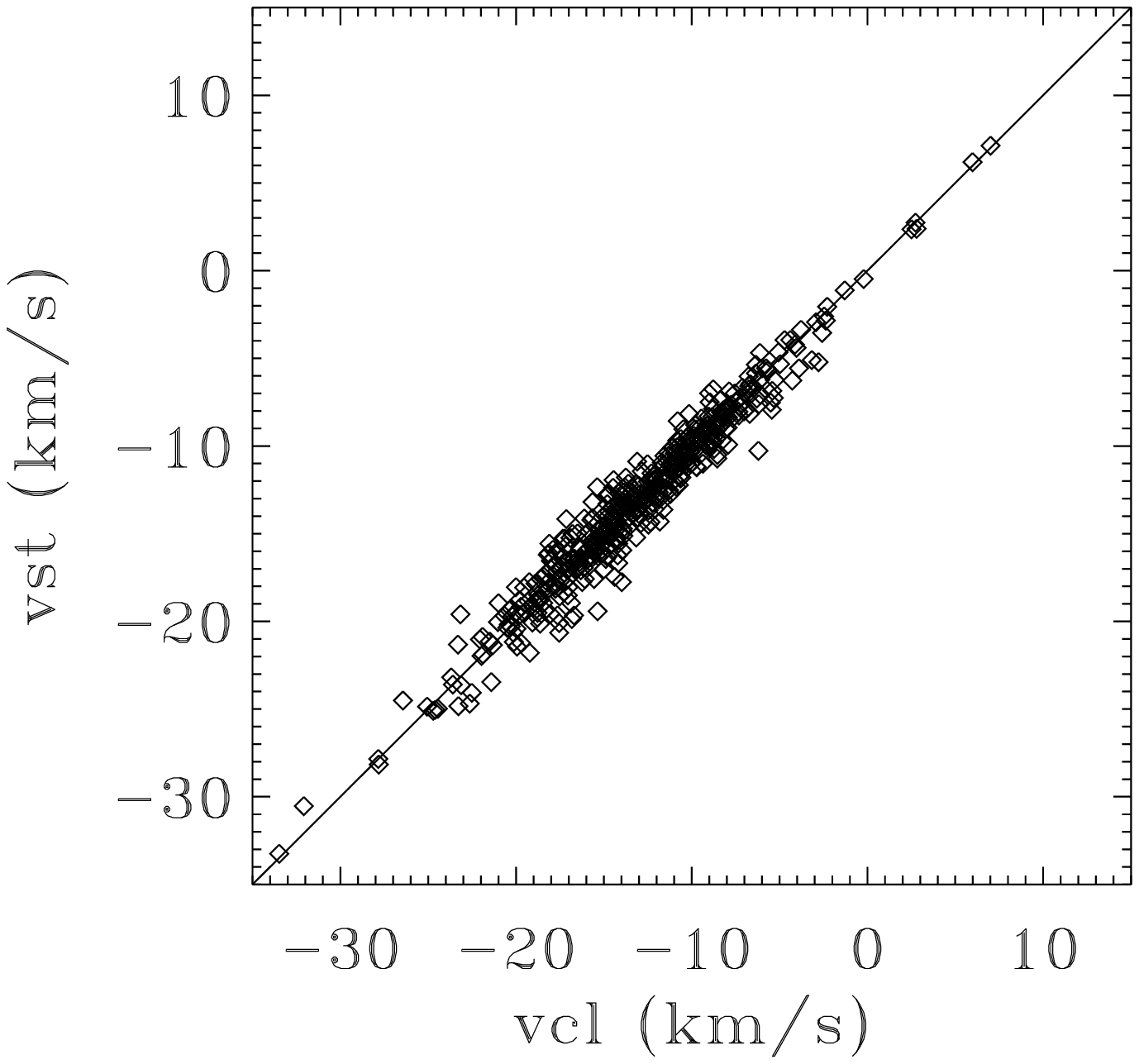}{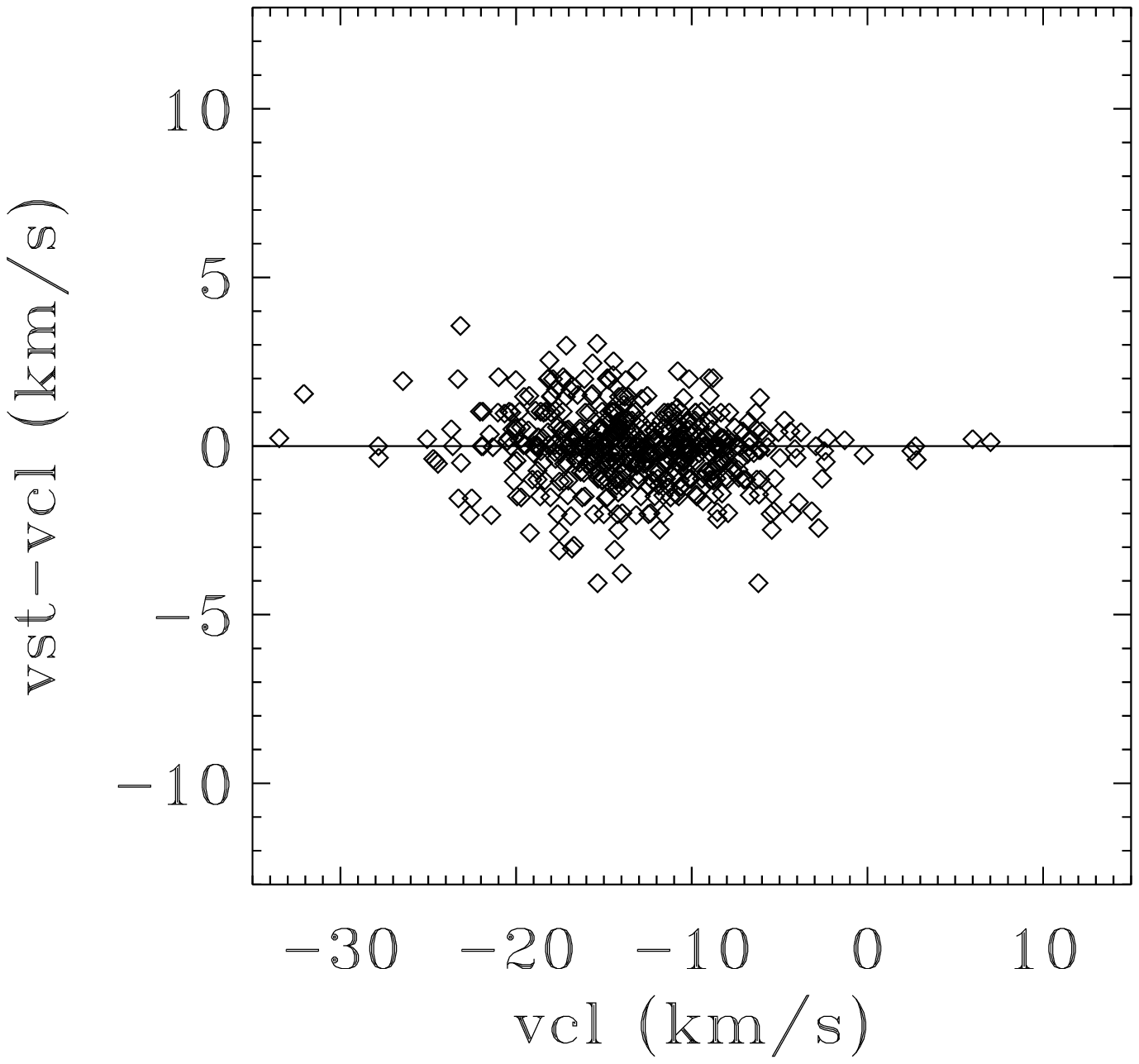}
\caption{Comparison of slitless vs slit (classical) radial 
velocities across Abell 35 for Chip 2 of FOCAS. 
}
\end{figure}

\begin{figure}
\figurenum{4}
\epsscale{1.0}
\plottwo{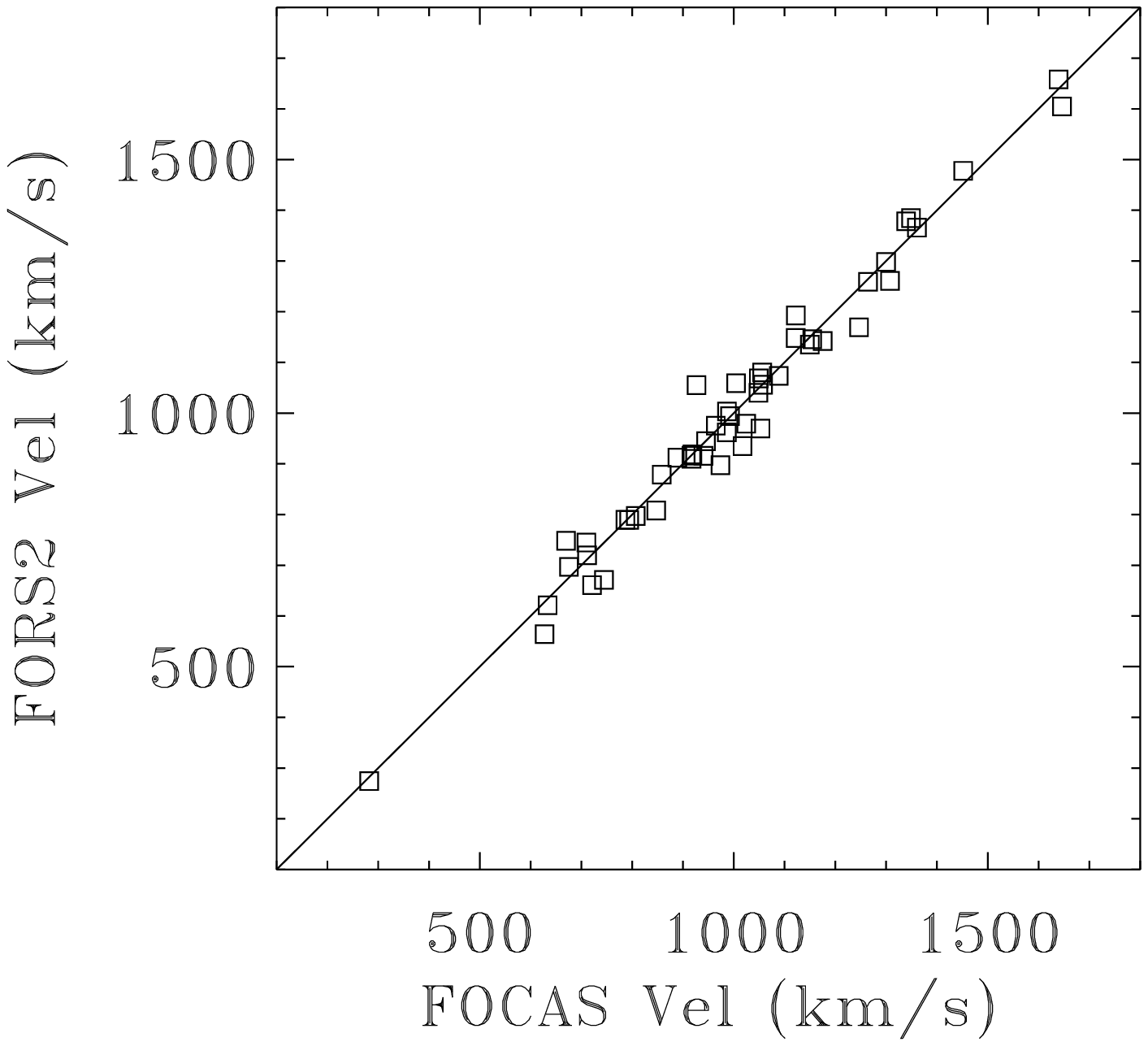}{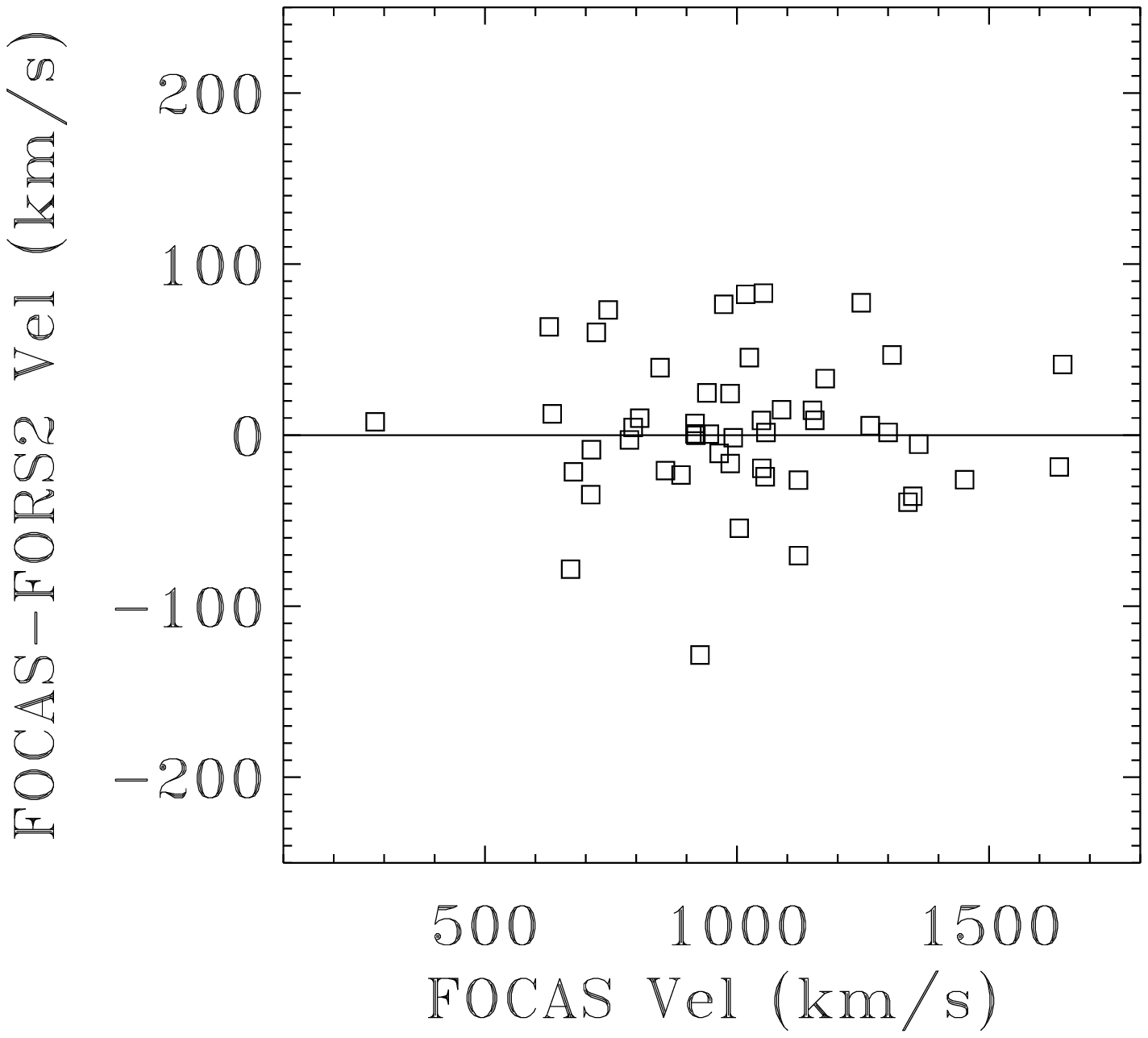}
\caption{Comparison of heliocentric slitless velocities 
measured with both FORS2 and FOCAS for 50 PN candidates. 
We plot FORS2 velocities (left) and (FOCAS $-$ FORS2) 
velocities (right) as a function of FOCAS velocities.
We obtain good agreement, with a standard deviation of 40 
km s$^{-1}$.
}
\end{figure}

\begin{figure}
\figurenum{5}
\epsscale{1.0}
\plotone{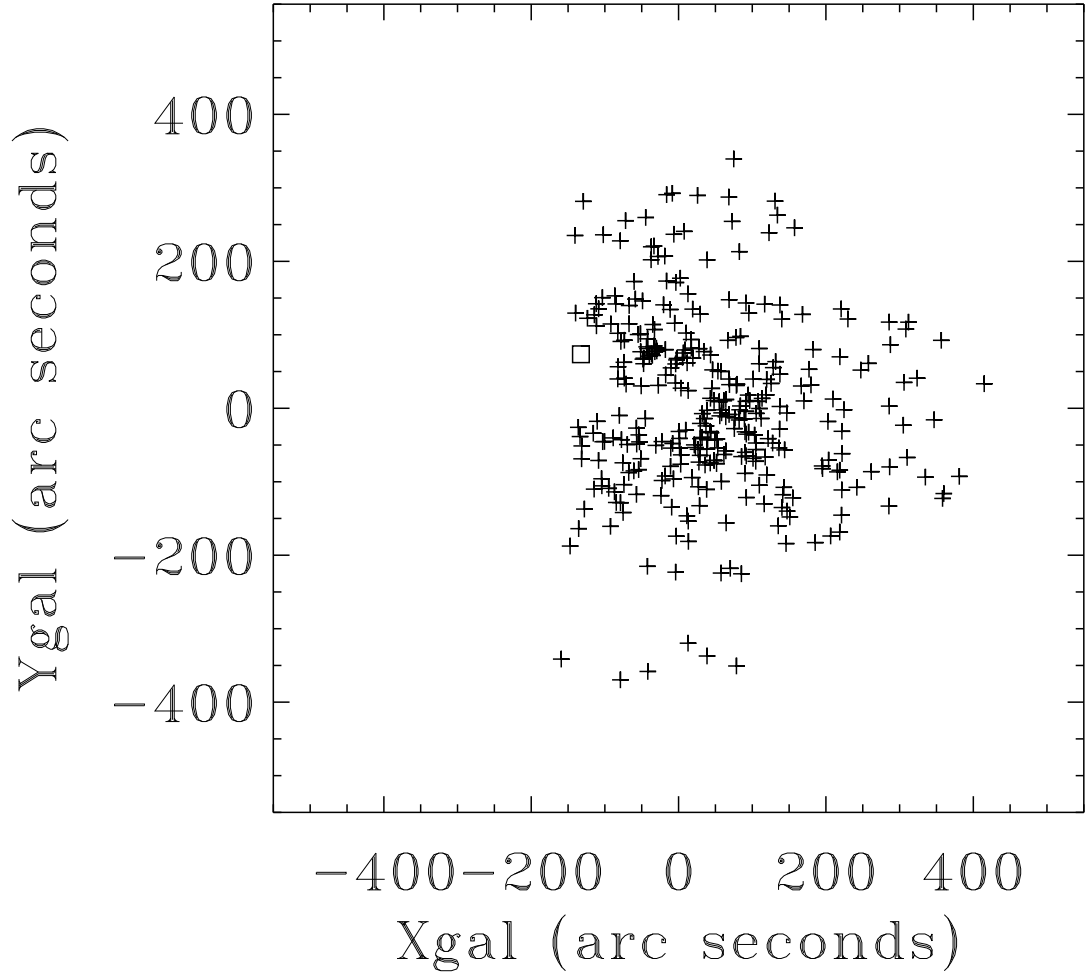}
\caption{Positions $x, y$ in arc seconds, relative to the 
center of M 60, for the 326 PN candidates with measured 
velocities (plus signs). The square marks the position of 
the center of NGC 4647. The $x$ coordinate runs in the direction 
of increasing RA, along the major axis of M 60. The $y$ 
coordinate runs along the minor axis, in the direction of 
increasing Declination. Because of these choices, the sky 
appears flipped from left to right.
}
\end{figure}

\begin{figure}
\figurenum{6}
\epsscale{1.0}
\plottwo{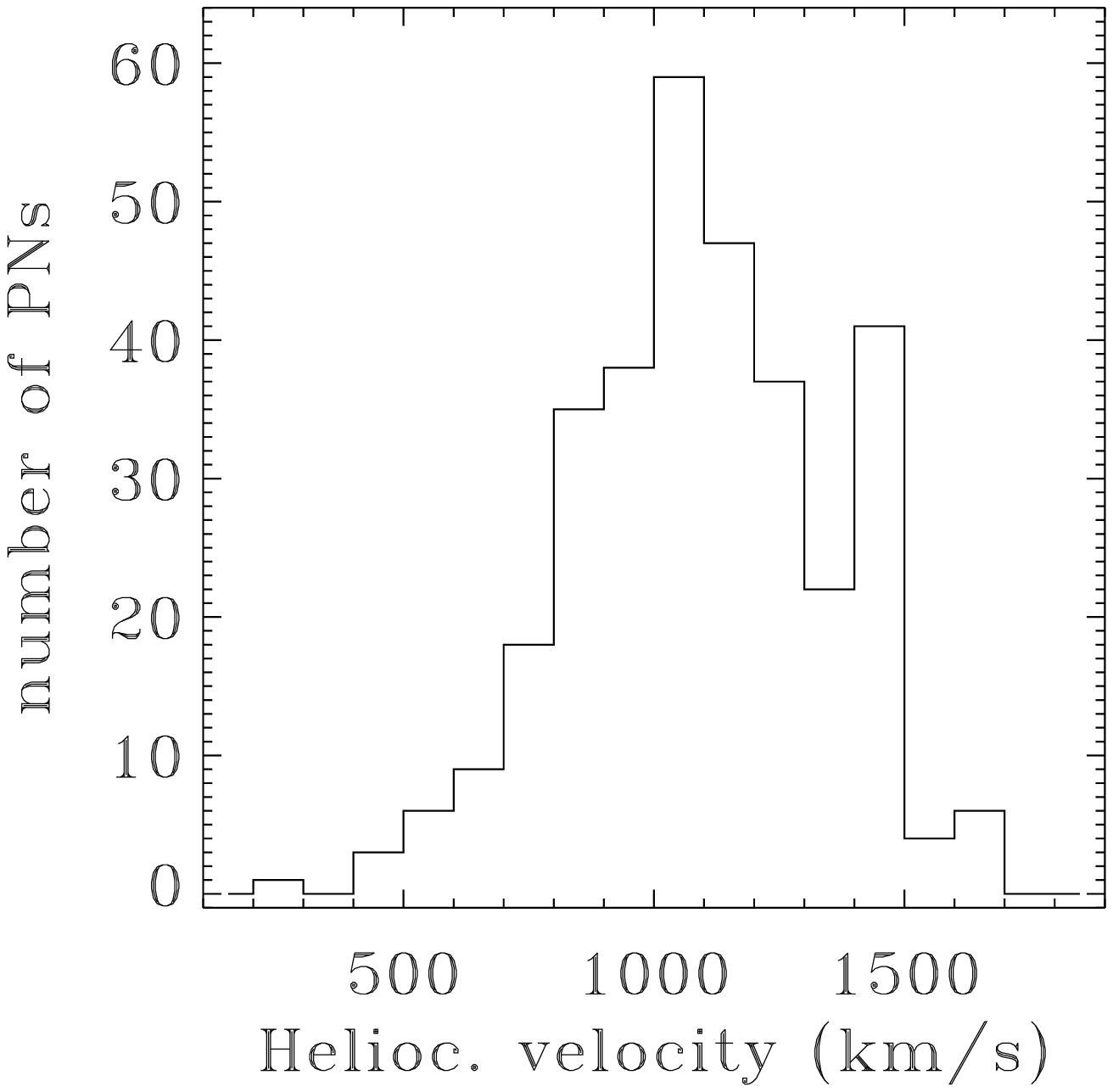}{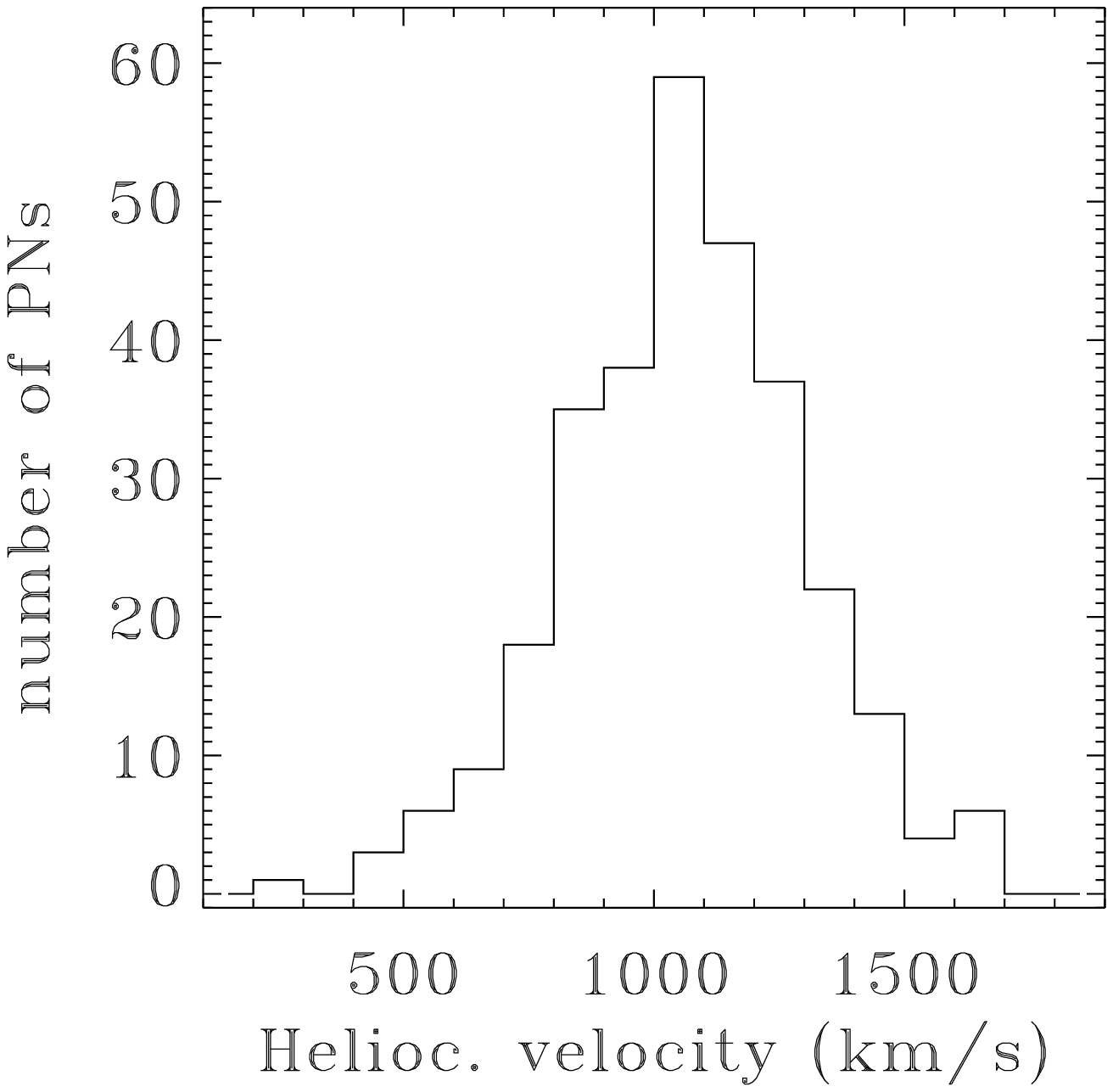}
\caption{{\it Left}: Number of PNs as a function of velocity 
for 326 detections. There is a peak at the velocity of NGC 4647
(1409 km s$^{-1}$), indicating some contamination. 
{\it Right}: Number of PNs as a function of velocity for the 
298 detections assigned to M 60 (see the text). 
The NGC 4647 peak has disappeared. 
}
\end{figure}

\begin{figure}
\figurenum{7}
\epsscale{1.0}
\plotone{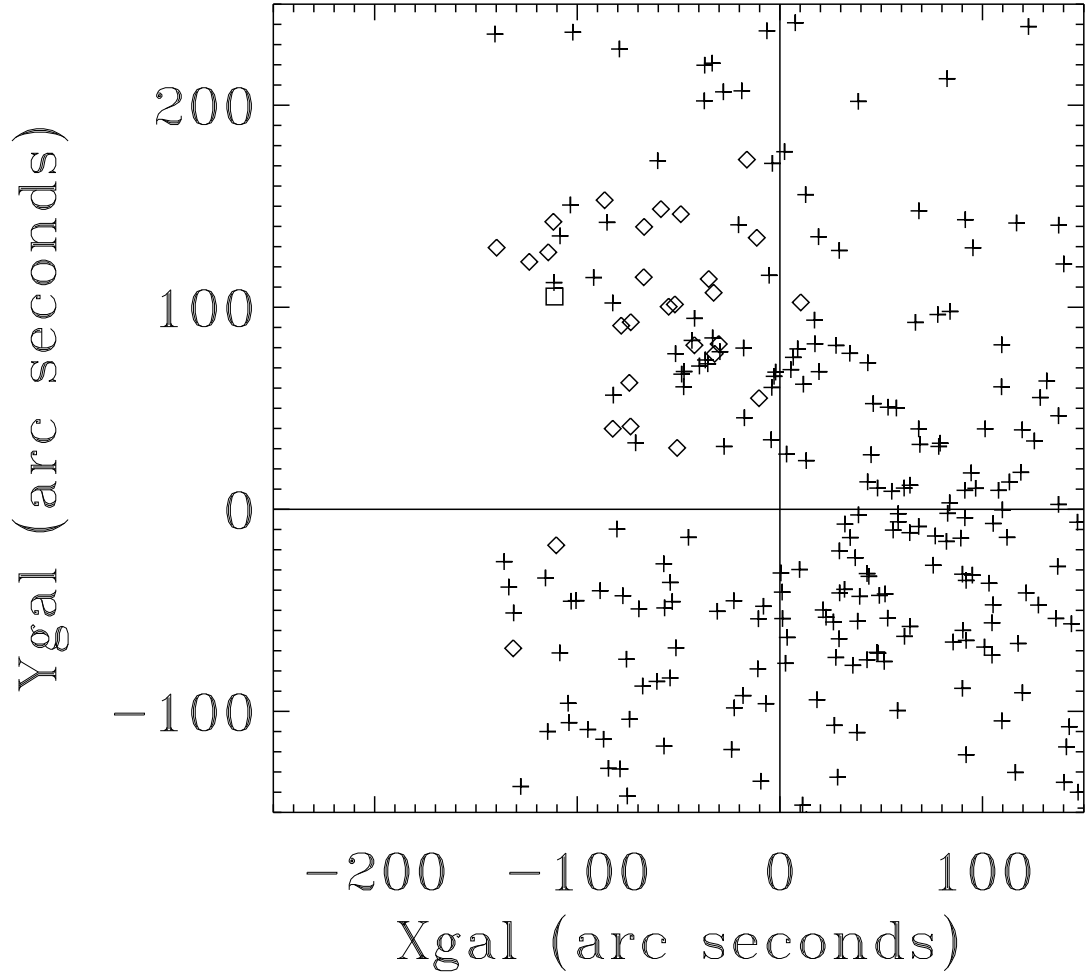}
\caption{Positions $x, y$ in arc seconds, relative to the 
center of M 60, for PNs near NGC 4647. The square
indicates the center of NGC 4647. Objects rejected from the total sample, 
because they may belong to NGC 4647, are indicated with diamonds. Plus
signs represent the M 60 PN sample.
}
\end{figure}

\begin{figure}
\figurenum{8}
\epsscale{1.0}
\plotone{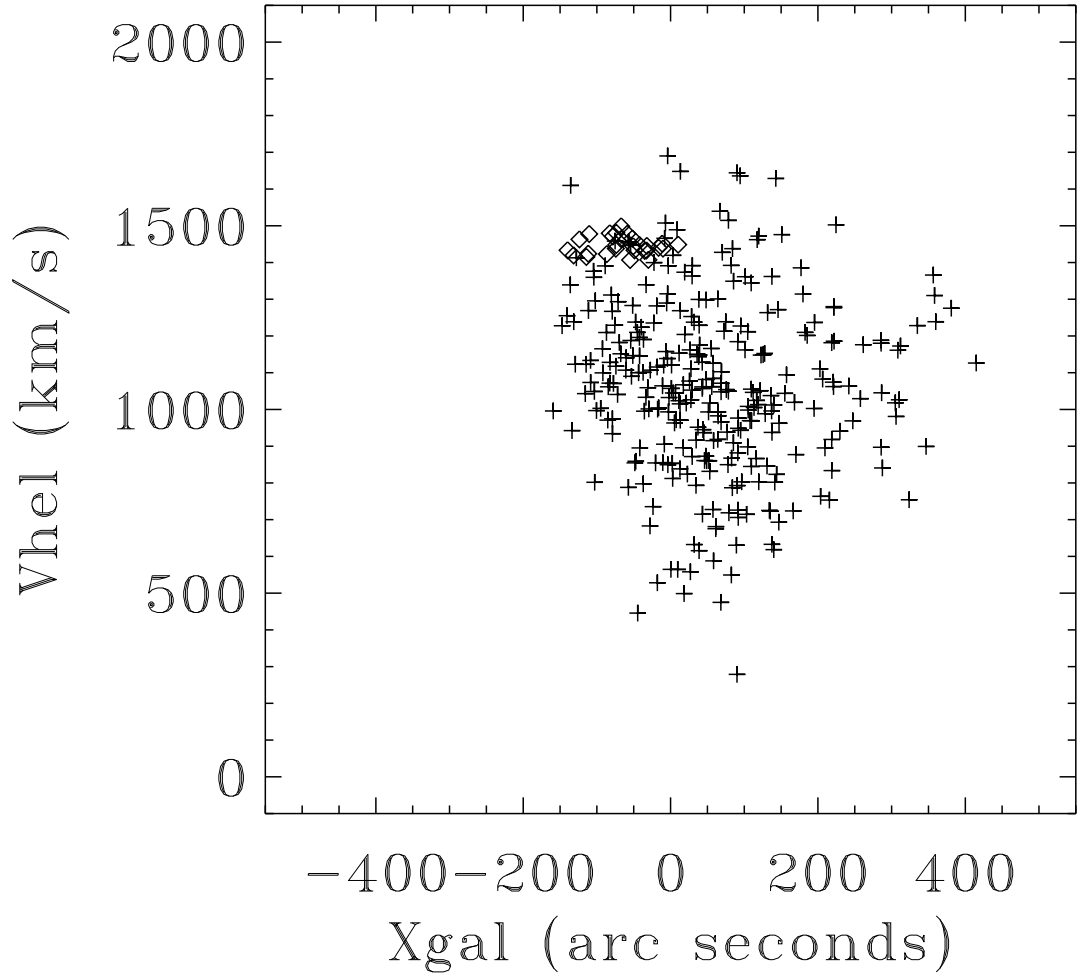}
\caption{Plus signs: heliocentric radial velocities of the 
298 PNs as a function of their $x$-coordinates in arc seconds 
relative to the center of M 60. Diamonds: 28 objects listed in Table 4.
}
\end{figure}

\begin{figure}
\figurenum{9}
\epsscale{1.0}
\plotone{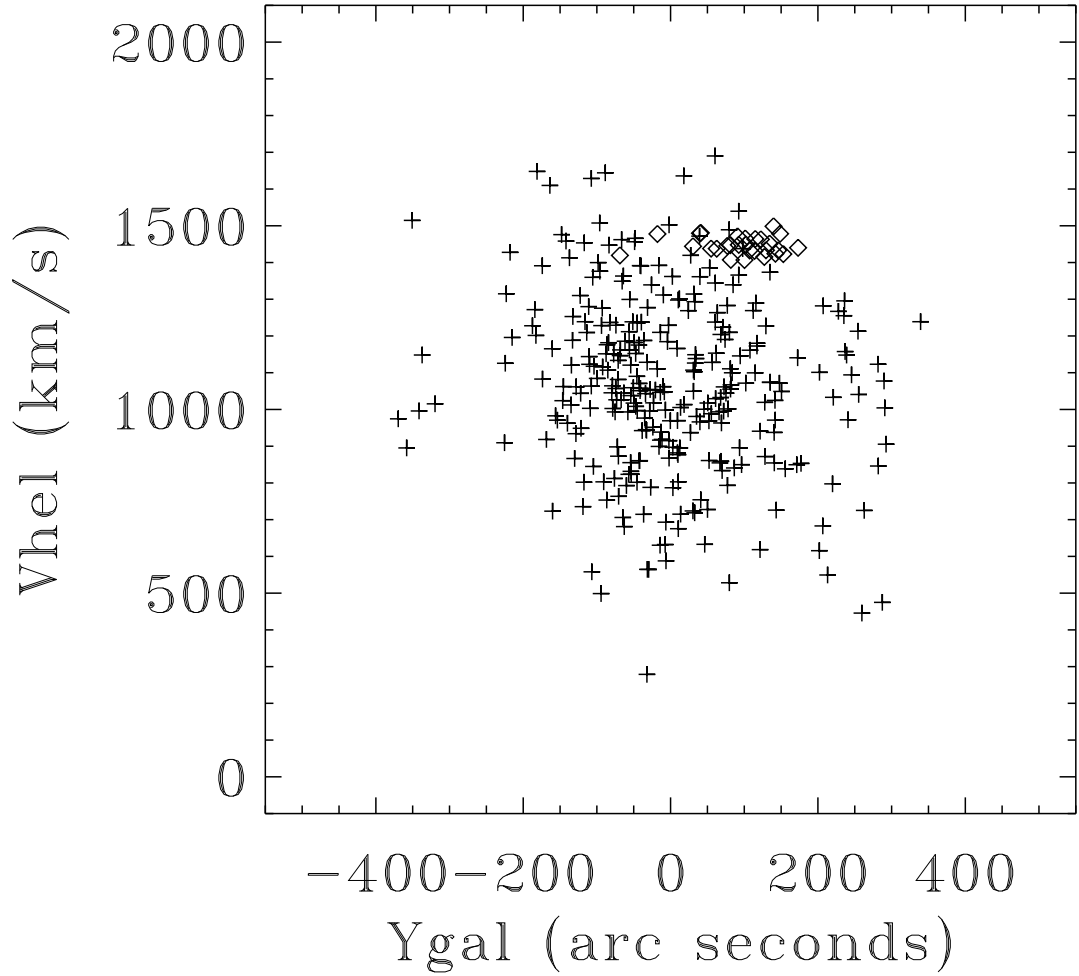}
\caption{Plus signs: heliocentric radial velocities of the 
298 PNs as a function of their $y$-coordinates in arc seconds 
relative to the center of M 60. Diamonds: 28 objects listed in Table 4.
}
\end{figure}

\begin{figure}
\figurenum{10}
\epsscale{1.0}
\plotone{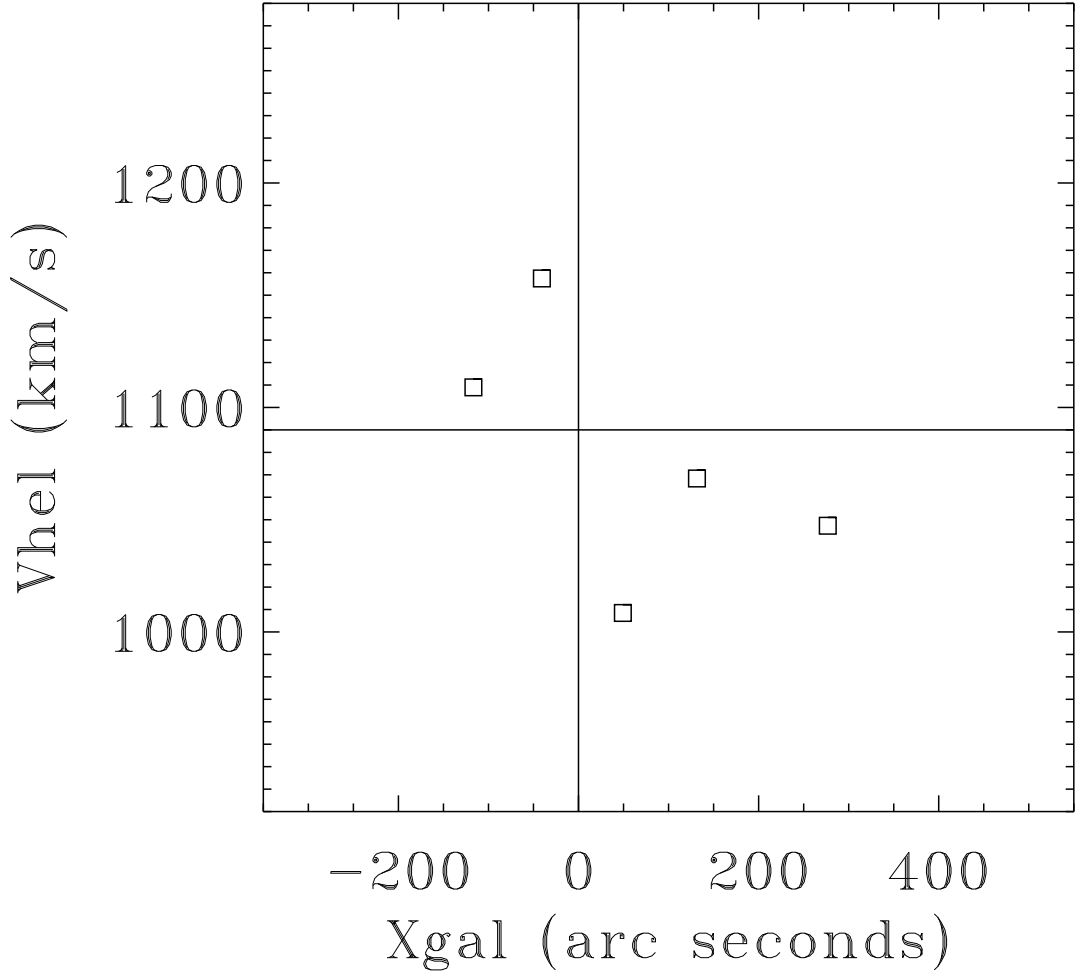}
\caption{Average velocities plotted as a function of average 
$x$ coordinate for five PN groups, defined in the text. The numbers of 
objects in each group, from left to right, are 8, 42, 84, 35 and 24. 
}
\end{figure}
\begin{figure}
\figurenum{11}
\epsscale{1.0}
\plotone{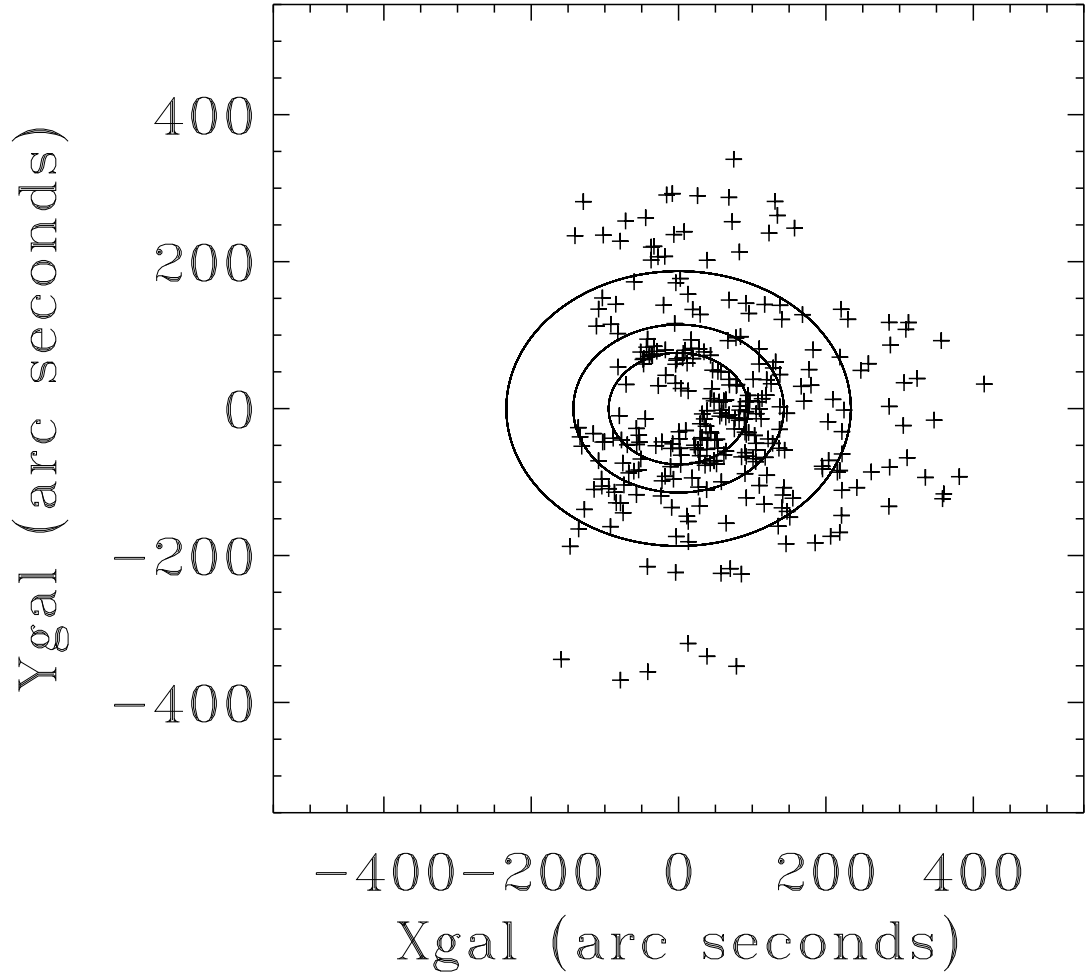}
\caption{The elliptical annuli used to divide the plane of 
the sky into four regions, for the calculation of LOSV$\sigma$s. 
Plus signs represent the 298 PNs in the M 60 sample.
}
\end{figure}

\begin{figure}
\figurenum{12}
\epsscale{1.0}
\plotone{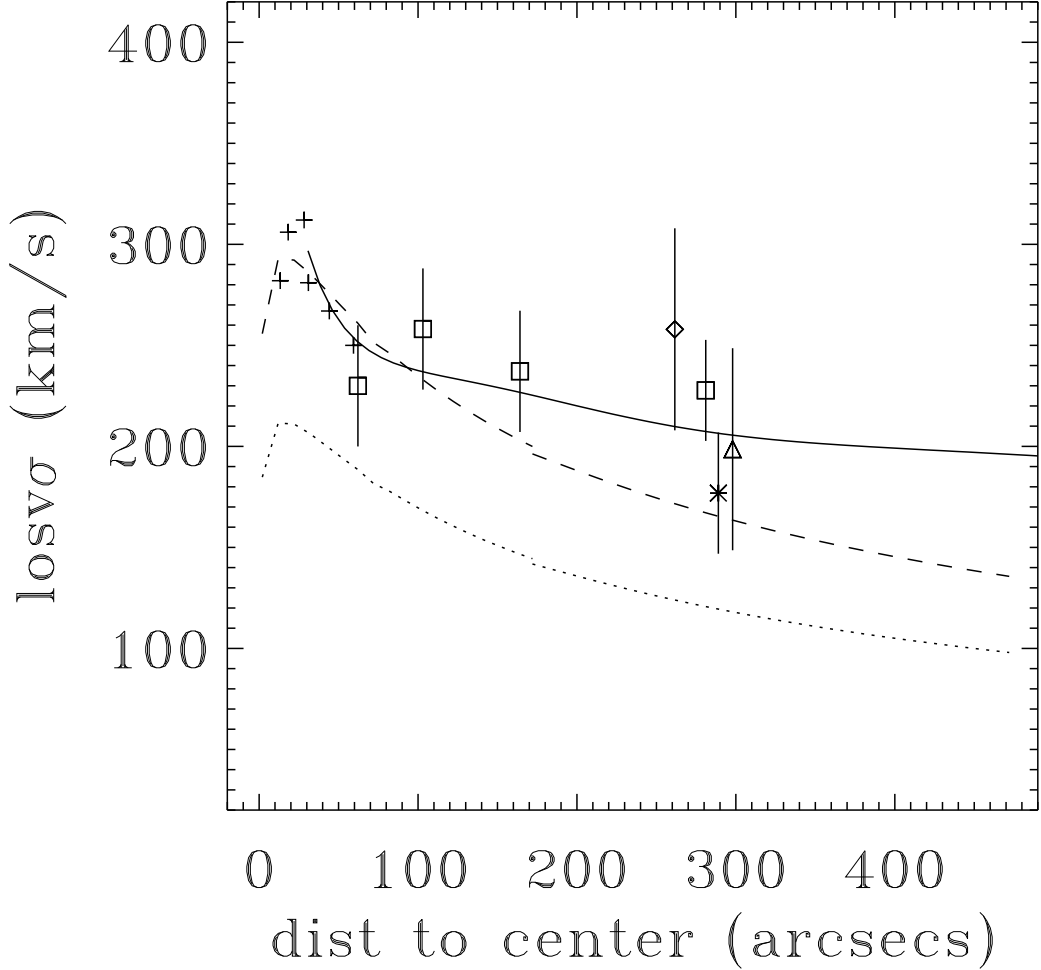}
\caption{LOSV$\sigma$ plotted as a function of 
average angular distance to the center of M 60.
The PNs were divided into 4 regions, as explained in the text.
These 4 data points are represented as squares. 
In addition, there is a diamond for PNs with large positive $y$,
a triangle for PNs with large negative $y$, and an asterisk for PNs
with large positive $x$. Plus signs are major axis, long-slit 
absorption-line data (Fisher et al. 1995, Pinkney et al. 2003). 
The dotted line represents the analytical model of 
Hernquist (1990), with a constant $M/L$ ratio, a total mass of 
6 $\times$ 10$^{11}$ $M_{\odot}$, and $R_{\rm e}$= 128$''$.
The dashed line is the same kind of model, but with a higher mass of
1.15 $\times$ 10$^{12}$ $M_{\odot}$. The solid line is a two-component
Hernquist mass distribution, as described in the text.
}
\end{figure}

\begin{figure}
\figurenum{13}
\epsscale{1.0}
\plotone{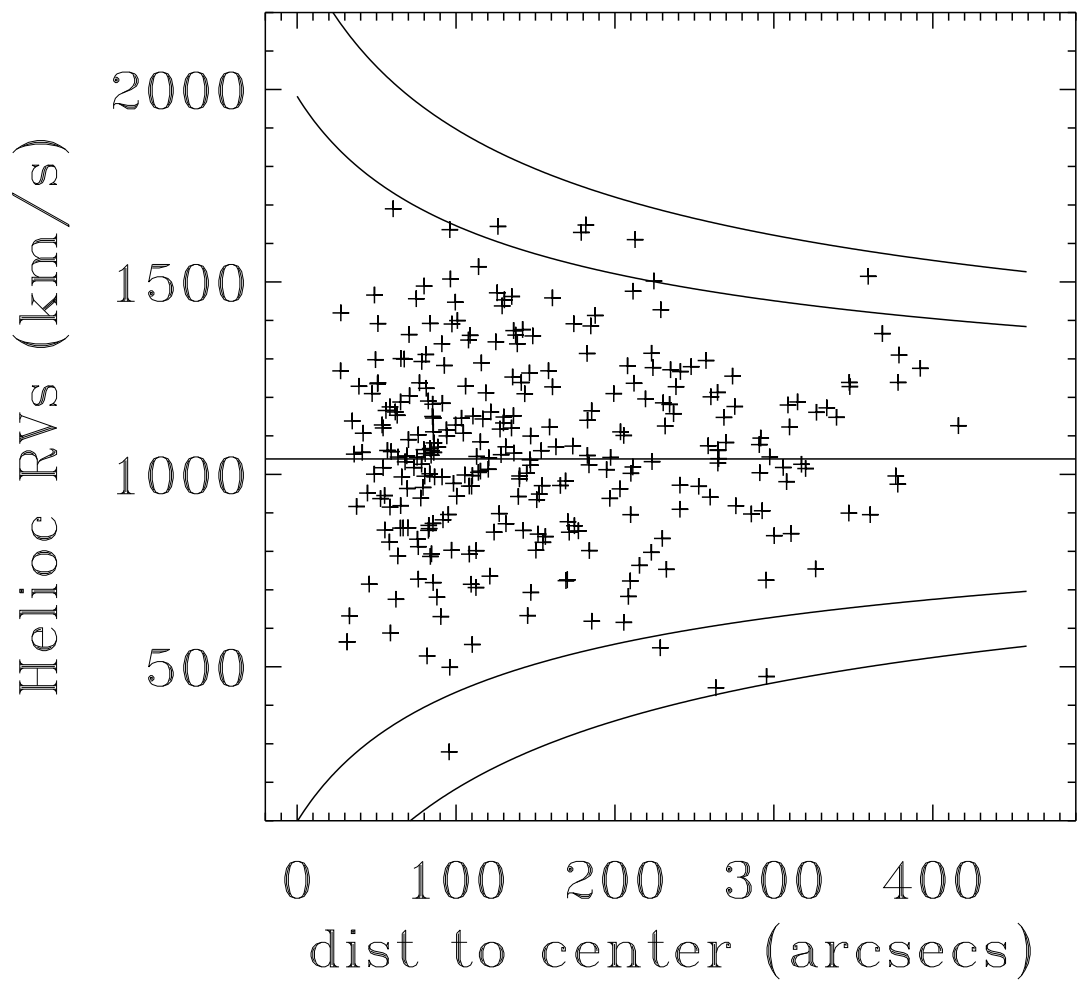}
\caption{Individual PN radial velocities plotted as a 
function of angular distance from the center of M 60. The solid 
lines are escape velocities for Hernquist models with total masses
1.2 x 10$^{12}$ $M_{\odot}$ (outer lines) and 6 x 10$^{11}$ $M_{\odot}$ 
(inner lines).
}
\end{figure}

\begin{figure}
\figurenum{14}
\epsscale{1.0}
\plotone{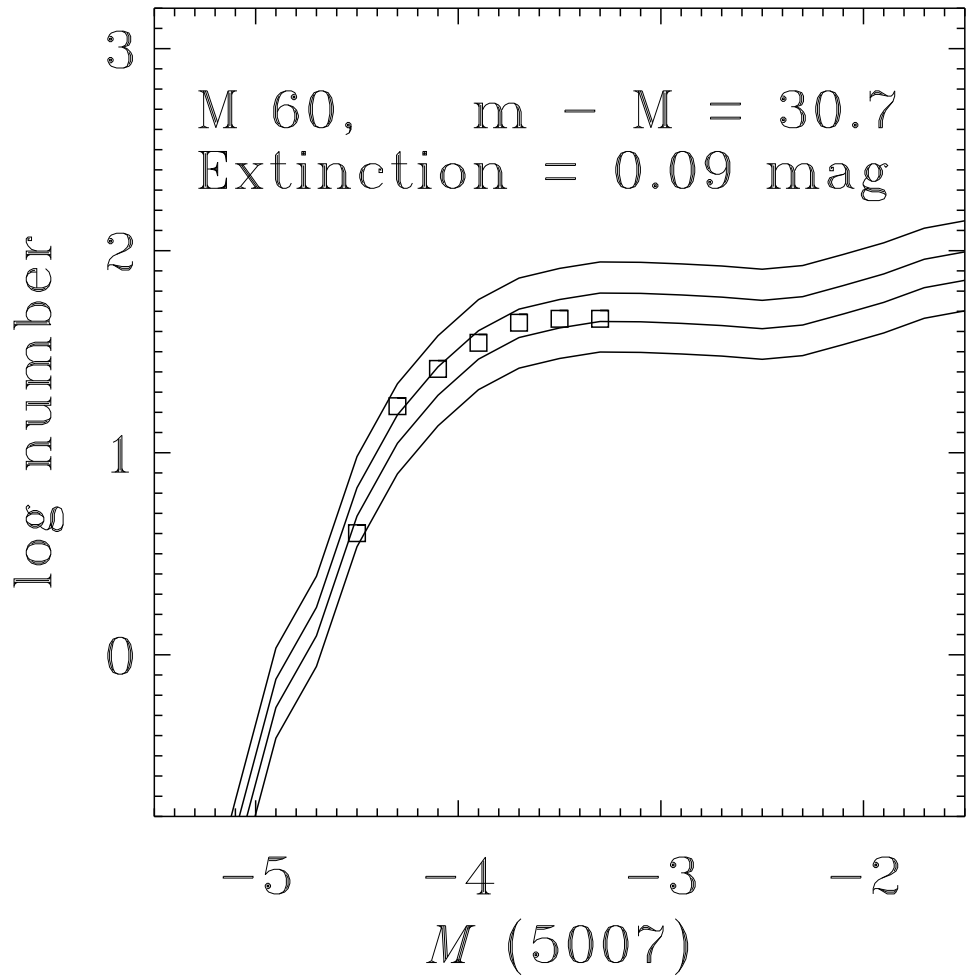}
\caption{Observed [O III] $\lambda$5007 PNLF of M 60 
(squares), with the statistically complete sample of 218 PNs binned 
into 0.2 mag intervals. The apparent magnitudes m(5007) have been 
transformed into absolute magnitudes M(5007) by adopting an extinction 
correction of 0.09 mag and a distance modulus $m - M$ = 30.7. 
The four lines are PNLF simulations (M\'endez and Soffner 1997) for 
four different total PN population sizes: 2400, 3400, 4700, and 6700 PNs. 
We estimate the best-fit sample size to be 4000. From this sample size 
it is possible to estimate the PN formation rate (see the text).
}
\end{figure}


\clearpage

\begin{deluxetable}{lrcrc}
\tablecaption{FORS2 observations and calibrations \label{tbl-1}}
\tablewidth{0pt}
\tablehead{
\colhead{FORS2 Field} & \colhead{Configuration} &
\colhead{FORS2 CCD Frame Ident (Chip 1)} &
\colhead{Exp (s)} & \colhead{Air Mass\tablenotemark{a}}}
\startdata
 Abell 35 + mask    &  On+grism  &  FORS2.2007-06-09T23:15:19.194  &  300  &  1.03 \\
 Abell 35 + mask    &   On-band  &  FORS2.2007-06-09T23:24:38.708  &  240  &  1.03 \\
 M 60 Field 2       &  Off-band  &  FORS2.2007-06-09T23:48:15.900  &  600  &  1.24 \\
 M 60 Field 2       &   On-band  &  FORS2.2007-06-09T23:59:17.372  & 1700  &  1.24 \\
 M 60 Field 2       &  On+grism  &  FORS2.2007-06-10T00:29:03.342  & 2100  &  1.24 \\
 M 60 Field 2       &  Off-band  &  FORS2.2007-06-10T01:05:46.856  &  600  &  1.27 \\
 M 60 Field 2       &   On-band  &  FORS2.2007-06-10T01:16:46.678  & 1700  &  1.30 \\
 M 60 Field 2       &  On+grism  &  FORS2.2007-06-10T01:46:34.724  & 2100  &  1.36 \\
 M 60 Field 2       &  Off-band  &  FORS2.2007-06-10T02:23:19.535  &  600  &  1.49 \\
 M 60 Field 2       &   On-band  &  FORS2.2007-06-10T02:34:18.977  & 2000  &  1.55 \\
 M 60 Field 2       &  On+grism  &  FORS2.2007-06-10T03:09:03.997  & 2400  &  1.79 \\
 Abell 35 + mask    &   On-band  &  FORS2.2007-06-10T22:56:45.928  &  240  &  1.06 \\
 Abell 35 + mask    &  On+grism  &  FORS2.2007-06-10T23:02:44.136  &  300  &  1.05 \\
 M 60 Field 1       &  Off-band  &  FORS2.2007-06-10T23:29:38.634  &  600  &  1.26 \\
 M 60 Field 1       &   On-band  &  FORS2.2007-06-10T23:40:54.568  & 1700  &  1.25 \\
 M 60 Field 1       &  On+grism  &  FORS2.2007-06-11T00:10:52.768  & 2100  &  1.24 \\
 M 60 Field 1       &  Off-band  &  FORS2.2007-06-11T00:47:48.811  &  600  &  1.26 \\
 M 60 Field 1       &   On-band  &  FORS2.2007-06-11T00:58:59.683  & 1700  &  1.27 \\
 M 60 Field 1       &  On+grism  &  FORS2.2007-06-11T01:28:57.702  & 2100  &  1.30 \\
 M 60 Field 1       &  Off-band  &  FORS2.2007-06-11T02:05:54.744  &  600  &  1.44 \\
 M 60 Field 1       &   On-band  &  FORS2.2007-06-11T02:17:07.606  & 1700  &  1.50 \\
 M 60 Field 1       &  On+grism  &  FORS2.2007-06-11T02:47:07.045  & 2100  &  1.65 \\
\enddata
\tablenotetext{a}{the air masses correspond to the middle of each exposure}
\end{deluxetable}

\clearpage

\begin{deluxetable}{lrcrc}
\tablecaption{FOCAS observations and calibrations \label{tbl-2}}
\tablewidth{0pt}
\tablehead{
\colhead{FOCAS Field} & \colhead{Configuration} &
\colhead{FOCAS Ident (Chip 1)} &
\colhead{Exp (s)} & \colhead{Air Mass\tablenotemark{a}}}
\startdata
 M 60 Field 1            &  Off-band  &  88415  &  220  &  1.01 \\
 M 60 Field 1            &   On-band  &  88417  & 2700  &  1.01 \\
 M 60 Field 1            &  On+grism  &  88419  & 2700  &  1.04 \\
 M 60 Field 1            &  On+grism  &  88421  & 2700  &  1.11 \\
 G 138 - 31              &   On-band  &  88429  &   60  &  1.02 \\
 G 138 - 31              &   On-band  &  88437  &  120  &  1.02 \\
 M 60 Field 1            &  Off-band  &  88523  &  220  &  1.56 \\
 M 60 Field 1            &   On-band  &  88525  & 2700  &  1.52 \\
 M 60 Field 1            &  On+grism  &  88527  & 2700  &  1.27 \\
 M 60 Field 1            &  On+grism  &  88529  & 2700  &  1.12 \\
 M 60 Field 1            &  Off-band  &  88535  &  220  &  1.04 \\
 M 60 Field 1            &   On-band  &  88537  & 2700  &  1.03 \\
 M 60 Field 1            &  On+grism  &  88539  & 2700  &  1.01 \\
 M 60 Field 1            &  On+grism  &  88541  & 2700  &  1.03 \\
 M 60 Field 1            &  Off-band  &  88547  &  220  &  1.11 \\
 M 60 Field 1            &   On-band  &  88549  & 2700  &  1.12 \\
 M 60 Field 1            &  On+grism  &  88551  & 2700  &  1.26 \\
 M 60 Field 1            &  Off-band  &  95465  &  220  &  1.01 \\
 M 60 Field 1            &   On-band  &  95469  & 2700  &  1.02 \\
 M 60 Field 1            &  On+grism  &  95471  & 2700  &  1.07 \\
 M 60 Field 1            &  On+grism  &  95473  & 2700  &  1.16 \\
 M 60 Field 1            &  On+grism  &  95479  & 2700  &  1.38 \\
 G 138 - 31              &   On-band  &  95481  &   60  &  1.02 \\
 G 138 - 31              &   On-band  &  95483  &   60  &  1.02 \\
 M 60 Field 3            &  Off-band  &  95581  &  220  &  1.27 \\
 M 60 Field 3            &   On-band  &  95583  & 2700  &  1.25 \\
 M 60 Field 3            &  On+grism  &  95585  & 2700  &  1.11 \\
 M 60 Field 3            &  On+grism  &  95587  & 2700  &  1.04 \\
 Abell 35 + mask         &   On-band  &  95597  &  300  &  1.36 \\
 Abell 35 + mask         &  On+grism  &  95599  &  600  &  1.36 \\
 M 60 Field 3            &  Off-band  &  95603  &  220  &  1.02 \\
 M 60 Field 3            &   On-band  &  95605  & 2700  &  1.03 \\
 M 60 Field 3            &  On+grism  &  95607  & 2700  &  1.08 \\
 M 60 Field 3            &  On+grism  &  95609  & 2200  &  1.20 \\
 M 60 Field 3            &  Off-band  &  95615  &  220  &  1.40 \\
 M 60 Field 3            &   On-band  &  95617  & 2700  &  1.43 \\
 G 138 - 31              &   On-band  &  95621  &   60  &  1.02 \\
 G 138 - 31              &   On-band  &  95623  &   60  &  1.02 \\
\enddata
\tablenotetext{a}{the air masses correspond to the middle of each exposure}
\end{deluxetable}

\clearpage

\begin{deluxetable}{rccrccrrrrr}
\tablecaption{PN Candidates in M 60\label{tbl-3}}
\tablewidth{0pt}
\rotate
\tabletypesize{\small}
\tablehead{

\colhead{ID} &
\multicolumn{3}{c}{$\alpha$} & \multicolumn{3}{c}{$\delta$} & 
\colhead{$x$, G} & \colhead{$y$, G} & \colhead{Helioc. RV} & \colhead{$m$} \\

\colhead{} & 
\multicolumn{3}{c}{(J2000)} & \multicolumn{3}{c}{(J2000)} & 
\colhead{} & \colhead{} & \colhead{(km s$^{-1}$)} & \colhead{(5007) }}

\startdata
      5 &   12 &   43 &   34.65 &   11 &   35 &   27.2 &   -112. &    112. &     1269. &    26.4 \\
      7 &   12 &   43 &   34.90 &   11 &   37 &   33.6 &   -141. &    235. &     1255. &    27.1 \\
     10 &   12 &   43 &   35.25 &   11 &   35 &   48.8 &   -108. &    135. &     1073. &    27.1 \\
     12 &   12 &   43 &   35.59 &   11 &   34 &   25.9 &    -82. &     56. &     1128. &    26.8 \\
     13 &   12 &   43 &   35.86 &   11 &   36 &    2.2 &   -103. &    151. &     1049. &    26.6 \\
     15 &   12 &   43 &   35.90 &   11 &   34 &    0.2 &    -71. &     33. &     1293. &    28.0 \\
     16 &   12 &   43 &   35.98 &   11 &   35 &   24.6 &    -92. &    115. &     1100. &    27.0 \\
     18 &   12 &   43 &   36.38 &   11 &   35 &   10.0 &    -82. &    102. &     1072. &    27.5 \\
     20 &   12 &   43 &   36.46 &   11 &   38 &   15.5 &   -129. &    282. &     1123. &    27.5 \\
     23 &   12 &   43 &   36.89 &   11 &   35 &   49.2 &    -85. &    142. &      971. &    27.1 \\
     26 &   12 &   43 &   37.44 &   11 &   37 &   24.5 &   -102. &    236. &     1296. &    27.9 \\
     28 &   12 &   43 &   37.94 &   11 &   34 &   20.8 &    -48. &     61. &     1238. &    26.9 \\
     29 &   12 &   43 &   37.97 &   11 &   34 &   37.7 &    -52. &     77. &     1283. &    26.8 \\
     30 &   12 &   43 &   37.99 &   11 &   34 &   27.2 &    -48. &     67. &      855. &    26.9 \\
     32 &   12 &   43 &   38.09 &   11 &   34 &   28.1 &    -47. &     68. &      859. &    27.5 \\
     35 &   12 &   43 &   38.62 &   11 &   34 &   42.0 &    -43. &     84. &     1100. &    27.5 \\
     36 &   12 &   43 &   38.64 &   11 &   34 &   28.7 &    -40. &     71. &     1224. &    26.7 \\
     38 &   12 &   43 &   38.74 &   11 &   33 &   47.2 &    -28. &     31. &     1107. &    26.3 \\
     40 &   12 &   43 &   38.81 &   11 &   37 &   10.5 &    -79. &    228. &     1267. &    27.2 \\
   2005 &   12 &   43 &   38.85 &   11 &   31 &   55.7 &      3. &    -76. &      812. &    -1.0 \\
     42 &   12 &   43 &   38.88 &   11 &   34 &   30.9 &    -37. &     74. &     1190. &    27.2 \\
     43 &   12 &   43 &   38.90 &   11 &   34 &   52.2 &    -42. &     94. &     1146. &    26.7 \\
     44 &   12 &   43 &   38.93 &   11 &   34 &   28.8 &    -36. &     72. &      995. &    26.8 \\
     45 &   12 &   43 &   39.07 &   11 &   36 &   12.2 &    -60. &    172. &     1141. &    27.3 \\
   2008 &   12 &   43 &   39.13 &   11 &   32 &   17.5 &      1. &    -54. &     1121. &    -1.0 \\
     48 &   12 &   43 &   39.31 &   11 &   34 &   40.5 &    -33. &     85. &     1339. &    27.6 \\
   2009 &   12 &   43 &   39.35 &   11 &   32 &   30.1 &      1. &    -41. &     1057. &    -1.0 \\
     53 &   12 &   43 &   39.43 &   11 &   34 &   33.0 &    -30. &     78. &     1002. &    27.3 \\
   2010 &   12 &   43 &   39.54 &   11 &   30 &   54.6 &     28. &   -133. &     1253. &    -1.0 \\
   2011 &   12 &   43 &   39.54 &   11 &   31 &   34.3 &     18. &    -94. &      499. &    -1.0 \\
   1090 &   12 &   43 &   39.48 &   11 &   32 &   39.5 &      0. &    -32. &      565. &    -1.0 \\
     59 &   12 &   43 &   39.65 &   11 &   33 &   58.3 &    -18. &     45. &     1000. &    27.0 \\
     63 &   12 &   43 &   39.77 &   11 &   37 &   35.2 &    -72. &    255. &     1041. &    27.0 \\
   2014 &   12 &   43 &   39.89 &   11 &   31 &   19.9 &     27. &   -107. &      558. &    -1.0 \\
     70 &   12 &   43 &   40.12 &   11 &   32 &   38.8 &     10. &    -30. &      565. &    27.0 \\
     71 &   12 &   43 &   40.23 &   11 &   34 &   31.7 &    -18. &     80. &      528. &    26.9 \\
     73 &   12 &   43 &   40.32 &   11 &   33 &   44.4 &     -4. &     34. &     1139. &    26.7 \\
     74 &   12 &   43 &   40.53 &   11 &   31 &   52.1 &     28. &    -73. &     1026. &    27.0 \\
     75 &   12 &   43 &   40.53 &   11 &   32 &   16.5 &     21. &    -50. &     1017. &    26.6 \\
     76 &   12 &   43 &   40.56 &   11 &   31 &   13.4 &     38. &   -110. &     1144. &    26.7 \\
     77 &   12 &   43 &   40.56 &   11 &   32 &   12.5 &     23. &    -53. &      824. &    28.0 \\
     78 &   12 &   43 &   40.70 &   11 &   33 &   35.5 &      3. &     27. &     1420. &    26.5 \\
     79 &   12 &   43 &   40.75 &   11 &   32 &    9.3 &     26. &    -56. &     1162. &    27.2 \\
     80 &   12 &   43 &   40.79 &   11 &   32 &    0.6 &     29. &    -64. &     1363. &    26.5 \\
     81 &   12 &   43 &   40.80 &   11 &   34 &    9.3 &     -4. &     60. &     1690. &    26.9 \\
     83 &   12 &   43 &   41.01 &   11 &   31 &   46.2 &     36. &    -77. &     1150. &    27.4 \\
     84 &   12 &   43 &   40.97 &   11 &   34 &   14.3 &     -3. &     66. &      993. &    27.5 \\
     85 &   12 &   43 &   41.06 &   11 &   34 &   16.1 &     -2. &     68. &     1032. &    27.3 \\
     86 &   12 &   43 &   41.11 &   11 &   36 &   34.9 &    -37. &    202. &     1101. &    28.0 \\
     87 &   12 &   43 &   41.14 &   11 &   35 &   31.3 &    -20. &    141. &      854. &    27.0 \\
     88 &   12 &   43 &   41.21 &   11 &   32 &   22.4 &     30. &    -41. &     1392. &    27.7 \\
     89 &   12 &   43 &   41.28 &   11 &   33 &   30.0 &     13. &     24. &     1269. &    26.7 \\
     92 &   12 &   43 &   41.40 &   11 &   32 &   23.5 &     32. &    -40. &     1238. &    27.4 \\
     93 &   12 &   43 &   41.43 &   11 &   36 &   51.9 &    -37. &    220. &      797. &    27.1 \\
   2021 &   12 &   43 &   41.50 &   11 &   30 &   22.6 &     65. &   -156. &      983. &    -1.0 \\
     95 &   12 &   43 &   41.51 &   11 &   31 &   47.0 &     43. &    -74. &     1057. &    27.4 \\
     96 &   12 &   43 &   41.55 &   11 &   32 &    6.7 &     38. &    -55. &     1299. &    27.2 \\
     97 &   12 &   43 &   41.57 &   11 &   32 &   42.5 &     29. &    -21. &     1053. &    -1.0 \\
     98 &   12 &   43 &   41.57 &   11 &   34 &   15.3 &      5. &     69. &      963. &    27.7 \\
    100 &   12 &   43 &   41.64 &   11 &   37 &   32.4 &    -45. &    260. &      446. &    26.9 \\
    101 &   12 &   43 &   41.69 &   11 &   36 &   51.9 &    -33. &    221. &     1033. &    27.9 \\
    102 &   12 &   43 &   41.69 &   11 &   35 &    3.2 &     -5. &    116. &     1290. &    26.6 \\
    103 &   12 &   43 &   41.76 &   11 &   34 &   20.9 &      7. &     75. &     1044. &    27.5 \\
    104 &   12 &   43 &   41.81 &   11 &   36 &   36.8 &    -28. &    207. &      683. &    27.8 \\
    105 &   12 &   43 &   41.83 &   11 &   32 &   18.2 &     39. &    -43. &     1175. &    27.2 \\
    106 &   12 &   43 &   41.85 &   11 &   34 &    6.8 &     12. &     62. &     1153. &    27.5 \\
    107 &   12 &   43 &   41.90 &   11 &   31 &   49.2 &     48. &    -71. &      873. &    -1.0 \\
    108 &   12 &   43 &   41.93 &   11 &   31 &   48.7 &     48. &    -71. &     1082. &    -1.0 \\
    110 &   12 &   43 &   41.98 &   11 &   32 &   54.6 &     32. &     -7. &      632. &    26.8 \\
    111 &   12 &   43 &   41.98 &   11 &   34 &   24.4 &      9. &     79. &     1489. &    27.1 \\
    112 &   12 &   43 &   42.02 &   11 &   32 &   37.2 &     37. &    -24. &      952. &    -1.0 \\
    113 &   12 &   43 &   42.04 &   11 &   32 &   47.5 &     35. &    -14. &      916. &    26.3 \\
    114 &   12 &   43 &   42.06 &   11 &   31 &   18.8 &     58. &   -100. &     1085. &    27.3 \\
    115 &   12 &   43 &   42.06 &   11 &   31 &   44.0 &     52. &    -75. &      993. &    26.9 \\
    116 &   12 &   43 &   42.27 &   11 &   32 &   28.2 &     43. &    -32. &     1128. &    -1.0 \\
    117 &   12 &   43 &   42.30 &   11 &   32 &   26.6 &     44. &    -33. &      945. &    26.7 \\
    118 &   12 &   43 &   42.41 &   11 &   36 &   34.9 &    -19. &    207. &     1282. &    27.9 \\
    119 &   12 &   43 &   42.47 &   11 &   34 &   10.8 &     19. &     68. &     1204. &    26.7 \\
    122 &   12 &   43 &   42.50 &   11 &   32 &   57.3 &     39. &     -3. &     1229. &    -1.0 \\
    124 &   12 &   43 &   42.58 &   11 &   34 &   24.6 &     17. &     82. &     1067. &    27.6 \\
    125 &   12 &   43 &   42.55 &   11 &   32 &    4.2 &     53. &    -54. &      832. &    26.7 \\
    126 &   12 &   43 &   42.67 &   11 &   32 &   16.1 &     52. &    -42. &      860. &    27.1 \\
    127 &   12 &   43 &   42.77 &   11 &   35 &   56.3 &     -4. &    171. &      850. &    26.8 \\
    128 &   12 &   43 &   42.77 &   11 &   34 &   36.0 &     17. &     94. &      896. &    27.5 \\
    129 &   12 &   43 &   42.94 &   11 &   31 &   53.3 &     62. &    -63. &      681. &    27.2 \\
    130 &   12 &   43 &   43.20 &   11 &   31 &   57.4 &     64. &    -58. &      993. &    26.7 \\
    131 &   12 &   43 &   43.25 &   11 &   34 &   21.2 &     28. &     81. &     1111. &    27.6 \\
    132 &   12 &   43 &   43.27 &   11 &   36 &    0.4 &      2. &    177. &      853. &    26.8 \\
    133 &   12 &   43 &   43.36 &   11 &   33 &    7.6 &     48. &     10. &     1298. &    27.1 \\
    134 &   12 &   43 &   43.43 &   11 &   33 &   24.4 &     45. &     27. &      937. &    26.5 \\
    136 &   12 &   43 &   43.49 &   11 &   32 &   45.6 &     56. &    -10. &     1062. &    -1.0 \\
    137 &   12 &   43 &   43.58 &   11 &   35 &   37.1 &     13. &    156. &      838. &    27.0 \\
    138 &   12 &   43 &   43.63 &   11 &   35 &   15.4 &     19. &    135. &     1374. &    27.8 \\
    139 &   12 &   43 &   43.73 &   11 &   32 &   48.9 &     58. &     -6. &      588. &    -1.0 \\
    140 &   12 &   43 &   43.75 &   11 &   37 &    0.4 &     -6. &    237. &     1158. &    27.6 \\
    141 &   12 &   43 &   43.79 &   11 &   33 &    4.3 &     55. &      9. &     1166. &    26.6 \\
    142 &   12 &   43 &   43.80 &   11 &   32 &   52.8 &     58. &     -2. &      915. &    26.5 \\
    143 &   12 &   43 &   43.90 &   11 &   30 &   48.9 &     92. &   -122. &      949. &    27.6 \\
    144 &   12 &   43 &   44.06 &   11 &   37 &   55.0 &    -16. &    291. &     1004. &    27.1 \\
    145 &   12 &   43 &   44.01 &   11 &   32 &   42.3 &     64. &    -12. &      919. &    26.4 \\
    146 &   12 &   43 &   44.13 &   11 &   34 &    8.9 &     43. &     72. &     1062. &    27.4 \\
    147 &   12 &   43 &   44.18 &   11 &   35 &    6.1 &     29. &    128. &      872. &    27.2 \\
    149 &   12 &   43 &   44.22 &   11 &   33 &    4.2 &     61. &     10. &      675. &    -1.0 \\
    150 &   12 &   43 &   44.36 &   11 &   31 &   21.2 &     90. &    -89. &     1644. &    27.7 \\
    151 &   12 &   43 &   44.35 &   11 &   32 &   44.1 &     68. &     -8. &     1048. &    27.7 \\
    152 &   12 &   43 &   44.40 &   11 &   33 &   45.0 &     53. &     50. &     1018. &    27.5 \\
    153 &   12 &   43 &   44.46 &   11 &   31 &   44.5 &     86. &    -66. &     1349. &    27.6 \\
    154 &   12 &   43 &   44.49 &   11 &   32 &   23.7 &     76. &    -28. &     1054. &    27.5 \\
    155 &   12 &   43 &   44.59 &   11 &   37 &   54.7 &     -8. &    292. &      906. &    27.7 \\
    156 &   12 &   43 &   44.66 &   11 &   33 &   43.5 &     58. &     50. &      728. &    27.3 \\
    157 &   12 &   43 &   44.74 &   11 &   37 &    0.6 &      8. &    241. &      972. &    27.1 \\
    158 &   12 &   43 &   44.81 &   11 &   32 &   37.4 &     77. &    -13. &      938. &    27.1 \\
    159 &   12 &   43 &   44.88 &   11 &   31 &   48.8 &     90. &    -60. &      793. &    27.7 \\
    160 &   12 &   43 &   44.90 &   11 &   31 &   43.6 &     92. &    -65. &      706. &    27.5 \\
    161 &   12 &   43 &   45.10 &   11 &   33 &   23.2 &     69. &     32. &     1103. &    27.1 \\
    162 &   12 &   43 &   45.34 &   11 &   30 &   34.2 &    116. &   -130. &      866. &    27.7 \\
    163 &   12 &   43 &   45.36 &   11 &   32 &   15.8 &     90. &    -32. &      279. &    27.0 \\
    164 &   12 &   43 &   45.41 &   11 &   32 &   46.6 &     83. &     -2. &      867. &    -1.0 \\
    165 &   12 &   43 &   45.42 &   11 &   32 &   12.3 &     92. &    -35. &      977. &    26.6 \\
    166 &   12 &   43 &   45.44 &   11 &   31 &   38.0 &    101. &    -68. &     1162. &    27.1 \\
    168 &   12 &   43 &   45.57 &   11 &   32 &   51.4 &     84. &      3. &      787. &    -1.0 \\
    169 &   12 &   43 &   45.61 &   11 &   31 &   33.2 &    105. &    -72. &      898. &    27.8 \\
    170 &   12 &   43 &   45.62 &   11 &   32 &   33.0 &     89. &    -14. &      630. &    27.0 \\
    171 &   12 &   43 &   45.67 &   11 &   32 &   14.0 &     95. &    -33. &      943. &    27.6 \\
    172 &   12 &   43 &   45.89 &   11 &   31 &   48.6 &    105. &    -56. &     1212. &    27.5 \\
    173 &   12 &   43 &   45.93 &   11 &   32 &   42.4 &     91. &     -4. &     1185. &    27.4 \\
    174 &   12 &   43 &   46.03 &   11 &   34 &   22.1 &     67. &     92. &     1540. &    27.5 \\
    175 &   12 &   43 &   46.07 &   11 &   30 &    0.4 &    135. &   -160. &      723. &    27.4 \\
    177 &   12 &   43 &   46.08 &   11 &   31 &   57.1 &    105. &    -47. &     1008. &    27.6 \\
    178 &   12 &   43 &   46.10 &   11 &   36 &   15.0 &     39. &    202. &      616. &    26.8 \\
    180 &   12 &   43 &   46.14 &   11 &   32 &    8.0 &    103. &    -37. &      714. &    27.0 \\
    181 &   12 &   43 &   46.17 &   11 &   32 &   55.5 &     91. &      9. &      882. &    -1.0 \\
    182 &   12 &   43 &   46.27 &   11 &   31 &   11.4 &    120. &    -91. &      803. &    27.7 \\
    183 &   12 &   43 &   46.35 &   11 &   29 &   34.3 &    146. &   -184. &     1272. &    26.9 \\
    185 &   12 &   43 &   46.52 &   11 &   33 &    3.0 &     94. &     18. &     1636. &    27.0 \\
    186 &   12 &   43 &   46.56 &   11 &   31 &   35.5 &    118. &    -66. &     1462. &    26.9 \\
    187 &   12 &   43 &   46.53 &   11 &   32 &   55.1 &     97. &     10. &      803. &    -1.0 \\
    189 &   12 &   43 &   46.82 &   11 &   34 &   23.0 &     78. &     96. &      850. &    27.6 \\
    190 &   12 &   43 &   46.80 &   11 &   37 &   43.1 &     26. &    290. &     1077. &    27.5 \\
    191 &   12 &   43 &   46.80 &   11 &   32 &   36.0 &    105. &     -7. &      998. &    -1.0 \\
    193 &   12 &   43 &   46.84 &   11 &   30 &   23.2 &    140. &   -135. &     1012. &    27.2 \\
    194 &   12 &   43 &   47.11 &   11 &   35 &   14.9 &     69. &    148. &     1072. &    27.3 \\
    196 &   12 &   43 &   47.21 &   11 &   30 &   16.7 &    147. &   -140. &      962. &    27.0 \\
    197 &   12 &   43 &   47.22 &   11 &   30 &   39.9 &    141. &   -118. &      802. &    27.6 \\
    198 &   12 &   43 &   47.13 &   11 &   32 &   27.6 &    112. &    -14. &     1046. &    26.8 \\
    199 &   12 &   43 &   47.21 &   11 &   32 &   41.3 &    110. &     -0. &      969. &    27.5 \\
    200 &   12 &   43 &   47.25 &   11 &   34 &   22.8 &     84. &     98. &     1438. &    27.2 \\
    201 &   12 &   43 &   47.26 &   11 &   31 &   58.7 &    122. &    -41. &     1051. &    27.4 \\
    202 &   12 &   43 &   47.26 &   11 &   32 &   51.2 &    108. &      9. &      970. &    27.9 \\
    204 &   12 &   43 &   47.32 &   11 &   30 &    8.1 &    151. &   -148. &     1476. &    27.5 \\
    205 &   12 &   43 &   47.49 &   11 &   30 &   49.2 &    143. &   -108. &     1629. &    27.4 \\
    206 &   12 &   43 &   47.55 &   11 &   31 &   51.2 &    128. &    -47. &     1152. &    27.1 \\
    208 &   12 &   43 &   47.69 &   11 &   32 &   53.7 &    113. &     14. &     1005. &    -1.0 \\
    209 &   12 &   43 &   48.00 &   11 &   31 &   42.7 &    136. &    -54. &     1038. &    27.2 \\
    210 &   12 &   43 &   48.04 &   11 &   30 &   32.1 &    155. &   -122. &     1044. &    27.2 \\
    212 &   12 &   43 &   48.14 &   11 &   32 &   56.9 &    119. &     18. &     1013. &    -1.0 \\
    214 &   12 &   43 &   48.26 &   11 &   33 &   40.2 &    109. &     61. &     1344. &    27.5 \\
    215 &   12 &   43 &   48.46 &   11 &   31 &   38.0 &    144. &    -57. &      823. &    28.0 \\
    216 &   12 &   43 &   48.54 &   11 &   35 &    4.8 &     92. &    143. &      726. &    26.6 \\
    217 &   12 &   43 &   48.52 &   11 &   32 &    7.2 &    137. &    -28. &      996. &    27.0 \\
    218 &   12 &   43 &   48.54 &   11 &   34 &   50.4 &     95. &    129. &     1227. &    27.4 \\
    219 &   12 &   43 &   48.55 &   11 &   33 &   17.0 &    120. &     39. &     1472. &    27.4 \\
    221 &   12 &   43 &   48.63 &   11 &   34 &    0.3 &    110. &     81. &     1055. &    28.2 \\
   1126 &   12 &   43 &   48.85 &   11 &   33 &   10.2 &    126. &     34. &     1149. &    -1.0 \\
    224 &   12 &   43 &   48.97 &   11 &   29 &   25.5 &    185. &   -183. &     1201. &    27.1 \\
    225 &   12 &   43 &   49.08 &   11 &   32 &   36.7 &    138. &      2. &     1362. &    27.3 \\
    226 &   12 &   43 &   49.18 &   11 &   36 &   14.5 &     82. &    213. &      549. &    28.0 \\
    227 &   12 &   43 &   49.25 &   11 &   36 &   57.0 &     73. &    254. &     1213. &    27.3 \\
    228 &   12 &   43 &   49.42 &   11 &   33 &   30.2 &    128. &     55. &      988. &    27.7 \\
    230 &   12 &   43 &   49.54 &   11 &   32 &   25.8 &    147. &     -6. &      693. &    26.3 \\
    231 &   12 &   43 &   49.54 &   11 &   37 &   30.0 &     68. &    287. &      475. &    27.3 \\
    232 &   12 &   43 &   49.78 &   11 &   33 &   37.3 &    132. &     64. &     1264. &    27.2 \\
    234 &   12 &   43 &   49.85 &   11 &   33 &   19.1 &    138. &     46. &      633. &    -1.0 \\
    235 &   12 &   43 &   50.17 &   11 &   34 &   56.7 &    117. &    142. &     1025. &    27.6 \\
    236 &   12 &   43 &   50.50 &   11 &   29 &   28.7 &    206. &   -174. &     1083. &    27.2 \\
    238 &   12 &   43 &   50.90 &   11 &   38 &   18.3 &     75. &    339. &     1239. &    27.1 \\
    239 &   12 &   43 &   51.36 &   11 &   32 &   35.6 &    170. &     10. &      877. &    -1.0 \\
    240 &   12 &   43 &   51.41 &   11 &   29 &   30.6 &    219. &   -168. &      918. &    27.5 \\
    241 &   12 &   43 &   51.40 &   11 &   31 &    0.1 &    195. &    -82. &     1237. &    27.3 \\
    242 &   12 &   43 &   51.47 &   11 &   32 &   56.2 &    166. &     30. &      724. &    27.2 \\
    243 &   12 &   43 &   51.42 &   11 &   31 &    3.8 &    195. &    -78. &     1003. &    26.8 \\
    244 &   12 &   43 &   51.52 &   11 &   34 &   50.1 &    138. &    141. &      938. &    27.3 \\
    246 &   12 &   43 &   51.98 &   11 &   29 &   52.2 &    221. &   -146. &     1062. &    27.6 \\
    248 &   12 &   43 &   52.14 &   11 &   31 &    9.3 &    204. &    -70. &      764. &    27.4 \\
    250 &   12 &   43 &   52.27 &   11 &   36 &   29.0 &    123. &    239. &     1148. &    27.7 \\
    251 &   12 &   43 &   52.37 &   11 &   32 &   54.1 &    180. &     32. &     1314. &    27.6 \\
    252 &   12 &   43 &   52.57 &   11 &   33 &   15.5 &    177. &     53. &     1386. &    27.0 \\
    253 &   12 &   43 &   52.61 &   11 &   30 &   25.2 &    222. &   -111. &     1280. &    27.4 \\
    254 &   12 &   43 &   52.65 &   11 &   30 &   50.6 &    216. &    -87. &      753. &    26.5 \\
    255 &   12 &   43 &   52.89 &   11 &   30 &   52.1 &    219. &    -84. &     1182. &    27.2 \\
    257 &   12 &   43 &   53.01 &   11 &   32 &    0.2 &    203. &    -18. &     1110. &    26.5 \\
    258 &   12 &   43 &   53.40 &   11 &   33 &   40.0 &    182. &     80. &     1210. &    27.8 \\
    259 &   12 &   43 &   53.45 &   11 &   36 &   49.1 &    134. &    263. &      725. &    26.7 \\
    260 &   12 &   43 &   53.49 &   11 &   31 &   12.9 &    222. &    -62. &     1186. &    27.4 \\
    261 &   12 &   43 &   53.57 &   11 &   37 &    8.4 &    131. &    282. &      846. &    26.8 \\
    262 &   12 &   43 &   54.00 &   11 &   30 &   23.6 &    242. &   -108. &     1064. &    26.8 \\
    263 &   12 &   43 &   53.99 &   11 &   32 &   27.9 &    210. &     12. &      895. &    26.8 \\
    264 &   12 &   43 &   54.02 &   11 &   31 &   42.4 &    222. &    -31. &     1277. &    28.0 \\
    266 &   12 &   43 &   54.67 &   11 &   36 &   26.5 &    157. &    246. &     1094. &    27.4 \\
    267 &   12 &   43 &   54.72 &   11 &   32 &   10.0 &    224. &     -2. &     1502. &    27.2 \\
    268 &   12 &   43 &   55.62 &   11 &   33 &   21.0 &    219. &     70. &      834. &    26.8 \\
    269 &   12 &   43 &   55.67 &   11 &   30 &   39.0 &    262. &    -86. &     1176. &    27.3 \\
    270 &   12 &   43 &   56.42 &   11 &   29 &   47.6 &    285. &   -133. &     1188. &    26.6 \\
    271 &   12 &   43 &   56.88 &   11 &   34 &   23.4 &    221. &    135. &     1074. &    27.5 \\
    273 &   12 &   43 &   57.17 &   11 &   32 &   56.2 &    247. &     52. &      969. &    27.6 \\
    274 &   12 &   43 &   57.24 &   11 &   34 &    7.9 &    230. &    122. &      941. &    26.3 \\
    275 &   12 &   43 &   57.43 &   11 &   30 &   38.8 &    286. &    -80. &     1045. &    28.0 \\
    276 &   12 &   43 &   58.01 &   11 &   33 &    2.3 &    258. &     61. &     1029. &    27.8 \\
    277 &   12 &   43 &   58.84 &   11 &   31 &   58.9 &    286. &      3. &      897. &    27.5 \\
    278 &   12 &   43 &   59.21 &   11 &   30 &   45.1 &    310. &    -67. &     1026. &    27.0 \\
    280 &   12 &   43 &   59.64 &   11 &   31 &   28.7 &    305. &    -23. &     1018. &    27.8 \\
    281 &   12 &   44 &    0.36 &   11 &   30 &   12.8 &    335. &    -94. &     1228. &    27.9 \\
    282 &   12 &   44 &    0.42 &   11 &   33 &   19.2 &    288. &     86. &      841. &    27.0 \\
    283 &   12 &   44 &    0.73 &   11 &   32 &   24.9 &    306. &     35. &      981. &    27.2 \\
    284 &   12 &   44 &    0.84 &   11 &   33 &   49.4 &    286. &    117. &     1180. &    27.5 \\
   1157 &   12 &   44 &    1.39 &   11 &   29 &   38.8 &    358. &   -123. &     1310. &    -1.0 \\
   1158 &   12 &   44 &    1.61 &   11 &   29 &   44.7 &    360. &   -116. &     1239. &    -1.0 \\
    287 &   12 &   44 &    1.99 &   11 &   32 &   25.7 &    324. &     41. &      754. &    27.0 \\
    288 &   12 &   44 &    2.17 &   11 &   33 &   34.3 &    308. &    108. &     1161. &    27.5 \\
    289 &   12 &   44 &    2.52 &   11 &   31 &   25.1 &    347. &    -16. &      899. &    27.7 \\
   1162 &   12 &   44 &    2.57 &   11 &   33 &   42.5 &    312. &    117. &     1173. &    -1.0 \\
    601 &   12 &   43 &   43.09 &   11 &   33 &   12.0 &     43. &     14. &      715. &    26.5 \\
    603 &   12 &   43 &   43.63 &   11 &   34 &   15.7 &     34. &     77. &      793. &    26.3 \\
    605 &   12 &   43 &   43.95 &   11 &   33 &   48.6 &     46. &     52. &      861. &    26.7 \\
    608 &   12 &   43 &   44.43 &   11 &   33 &    4.9 &     64. &     12. &     1301. &    26.8 \\
    610 &   12 &   43 &   45.20 &   11 &   33 &   30.8 &     68. &     40. &      966. &    26.7 \\
    611 &   12 &   43 &   45.70 &   11 &   33 &   19.8 &     78. &     31. &     1050. &    26.5 \\
    612 &   12 &   43 &   45.78 &   11 &   33 &   21.1 &     79. &     33. &      719. &    26.4 \\
    613 &   12 &   43 &   47.36 &   11 &   33 &   22.2 &    101. &     40. &     1361. &    26.8 \\
    615 &   12 &   43 &   51.35 &   11 &   34 &   31.0 &    140. &    121. &      618. &    27.4 \\
    617 &   12 &   43 &   53.30 &   11 &   34 &   29.9 &    168. &    128. &     1019. &    27.4 \\
   1005 &   12 &   43 &   23.52 &   11 &   28 &   21.4 &   -159. &   -341. &      996. &    -1.0 \\
   1016 &   12 &   43 &   27.02 &   11 &   30 &   47.0 &   -147. &   -188. &     1228. &    -1.0 \\
   1022 &   12 &   43 &   28.20 &   11 &   31 &    6.9 &   -136. &   -164. &     1610. &    -1.0 \\
   1023 &   12 &   43 &   28.30 &   11 &   27 &   33.4 &    -79. &   -370. &      974. &    -1.0 \\
   1026 &   12 &   43 &   29.18 &   11 &   31 &   30.6 &   -128. &   -137. &     1413. &    -1.0 \\
   1033 &   12 &   43 &   30.46 &   11 &   32 &   54.5 &   -131. &    -51. &     1239. &    -1.0 \\
   1034 &   12 &   43 &   30.53 &   11 &   31 &   53.4 &   -115. &   -110. &     1123. &    -1.0 \\
   1035 &   12 &   43 &   30.53 &   11 &   33 &    7.4 &   -134. &    -38. &      942. &    -1.0 \\
   1036 &   12 &   43 &   30.60 &   11 &   33 &   20.2 &   -136. &    -26. &     1339. &    -1.0 \\
   1037 &   12 &   43 &   30.94 &   11 &   27 &   34.8 &    -42. &   -358. &      895. &    -1.0 \\
   1038 &   12 &   43 &   31.10 &   11 &   30 &   58.7 &    -92. &   -161. &     1165. &    -1.0 \\
   1040 &   12 &   43 &   31.30 &   11 &   31 &   54.9 &   -104. &   -106. &     1360. &    -1.0 \\
   1043 &   12 &   43 &   31.44 &   11 &   32 &    4.3 &   -104. &    -96. &     1376. &    -1.0 \\
   1044 &   12 &   43 &   31.61 &   11 &   32 &   29.4 &   -109. &    -71. &     1134. &    -1.0 \\
   1045 &   12 &   43 &   31.80 &   11 &   33 &    7.1 &   -116. &    -34. &     1043. &    -1.0 \\
   1046 &   12 &   43 &   31.85 &   11 &   31 &   49.2 &    -95. &   -109. &     1003. &    -1.0 \\
   1047 &   12 &   43 &   32.18 &   11 &   31 &   28.1 &    -85. &   -128. &     1062. &    -1.0 \\
   1048 &   12 &   43 &   32.28 &   11 &   31 &   42.6 &    -87. &   -114. &     1209. &    -1.0 \\
   1049 &   12 &   43 &   32.43 &   11 &   32 &   52.7 &   -103. &    -46. &      802. &    -1.0 \\
   1052 &   12 &   43 &   32.55 &   11 &   31 &   12.4 &    -75. &   -142. &     1458. &    -1.0 \\
   1053 &   12 &   43 &   32.55 &   11 &   31 &   26.3 &    -79. &   -128. &      934. &    -1.0 \\
   1054 &   12 &   43 &   32.59 &   11 &   32 &   52.4 &   -101. &    -45. &      997. &    -1.0 \\
   1056 &   12 &   43 &   33.29 &   11 &   31 &   48.8 &    -74. &   -104. &     1118. &    -1.0 \\
   1057 &   12 &   43 &   33.43 &   11 &   29 &   53.2 &    -42. &   -215. &     1196. &    -1.0 \\
   1058 &   12 &   43 &   33.46 &   11 &   32 &   54.0 &    -89. &    -40. &     1391. &    -1.0 \\
   1059 &   12 &   43 &   33.72 &   11 &   32 &   17.9 &    -76. &    -74. &     1229. &    -1.0 \\
   1060 &   12 &   43 &   34.01 &   11 &   32 &    3.0 &    -68. &    -88. &     1151. &    -1.0 \\
   1061 &   12 &   43 &   34.15 &   11 &   32 &   48.7 &    -78. &    -43. &     1071. &    -1.0 \\
   1062 &   12 &   43 &   34.18 &   11 &   31 &   31.6 &    -57. &   -117. &     1454. &    -1.0 \\
   1063 &   12 &   43 &   34.51 &   11 &   32 &    3.4 &    -61. &    -85. &     1107. &    -1.0 \\
   1064 &   12 &   43 &   34.54 &   11 &   33 &   21.3 &    -80. &    -10. &     1312. &    -1.0 \\
   1066 &   12 &   43 &   34.56 &   11 &   32 &   40.4 &    -70. &    -49. &     1182. &    -1.0 \\
   1067 &   12 &   43 &   34.97 &   11 &   32 &    3.4 &    -54. &    -84. &     1447. &    -1.0 \\
   1070 &   12 &   43 &   35.21 &   11 &   27 &   57.9 &     13. &   -320. &     1015. &    -1.0 \\
   1071 &   12 &   43 &   35.40 &   11 &   32 &   37.6 &    -57. &    -49. &     1456. &    -1.0 \\
   1072 &   12 &   43 &   35.42 &   11 &   32 &   17.0 &    -51. &    -69. &     1146. &    -1.0 \\
   1073 &   12 &   43 &   35.71 &   11 &   32 &   39.6 &    -53. &    -46. &     1090. &    -1.0 \\
   1074 &   12 &   43 &   35.76 &   11 &   32 &   58.8 &    -57. &    -27. &      788. &    -1.0 \\
   1075 &   12 &   43 &   35.81 &   11 &   29 &   35.6 &     -4. &   -223. &     1315. &    -1.0 \\
   1076 &   12 &   43 &   35.81 &   11 &   32 &   49.0 &    -54. &    -36. &     1188. &    -1.0 \\
   1077 &   12 &   43 &   36.34 &   11 &   31 &   21.3 &    -24. &   -119. &      735. &    -1.0 \\
   1078 &   12 &   43 &   36.60 &   11 &   27 &   34.3 &     39. &   -337. &     1148. &    -1.0 \\
   1079 &   12 &   43 &   36.79 &   11 &   31 &   40.9 &    -22. &    -98. &     1400. &    -1.0 \\
   1080 &   12 &   43 &   36.79 &   11 &   33 &    8.2 &    -45. &    -14. &     1210. &    -1.0 \\
   1081 &   12 &   43 &   37.01 &   11 &   31 &    2.4 &     -9. &   -135. &     1120. &    -1.0 \\
   1082 &   12 &   43 &   37.08 &   11 &   32 &   29.2 &    -31. &    -50. &     1059. &    -1.0 \\
   1083 &   12 &   43 &   37.18 &   11 &   31 &   45.7 &    -18. &    -92. &     1116. &    -1.0 \\
   1084 &   12 &   43 &   37.71 &   11 &   32 &   32.1 &    -23. &    -45. &     1235. &    -1.0 \\
   1085 &   12 &   43 &   37.85 &   11 &   31 &   38.9 &     -7. &    -96. &     1508. &    -1.0 \\
   1086 &   12 &   43 &   37.89 &   11 &   31 &   56.5 &    -11. &    -79. &     1064. &    -1.0 \\
   1087 &   12 &   43 &   38.35 &   11 &   32 &   20.4 &    -11. &    -54. &      855. &    -1.0 \\
   1088 &   12 &   43 &   38.62 &   11 &   32 &   25.7 &     -8. &    -48. &     1466. &    -1.0 \\
   1089 &   12 &   43 &   38.98 &   11 &   27 &   10.9 &     78. &   -351. &     1515. &    -1.0 \\
   1091 &   12 &   43 &   39.84 &   11 &   29 &   18.4 &     58. &   -224. &     1126. &    -1.0 \\
   1092 &   12 &   43 &   40.78 &   11 &   29 &   21.4 &     70. &   -218. &     1428. &    -1.0 \\
   1094 &   12 &   43 &   41.64 &   11 &   29 &   10.3 &     85. &   -225. &      909. &    -1.0 \\
   1163 &   12 &   44 &    3.41 &   11 &   30 &    1.9 &    381. &    -93. &     1276. &    -1.0 \\
   1164 &   12 &   44 &    5.06 &   11 &   33 &    7.1 &    356. &     92. &     1366. &    -1.0 \\
   1166 &   12 &   44 &    7.85 &   11 &   31 &   54.9 &    415. &     33. &     1126. &    -1.0 \\
   2001 &   12 &   43 &   36.72 &   11 &   30 &   22.7 &     -3. &   -174. &     1391. &    -1.0 \\
   2002 &   12 &   43 &   37.68 &   11 &   30 &   11.5 &     13. &   -181. &     1648. &    -1.0 \\
   2003 &   12 &   43 &   38.16 &   11 &   30 &   38.2 &     13. &   -154. &      971. &    -1.0 \\
   2004 &   12 &   43 &   38.16 &   11 &   30 &   45.7 &     11. &   -146. &     1024. &    -1.0 \\
   2007 &   12 &   43 &   39.12 &   11 &   32 &    7.8 &      4. &    -63. &     1045. &    -1.0 \\
   2029 &   12 &   43 &   42.48 &   11 &   32 &   16.1 &     49. &    -43. &      860. &    -1.0 \\
   2041 &   12 &   43 &   45.12 &   11 &   32 &   33.4 &     82. &    -16. &     1392. &    -1.0 \\
   2042 &   12 &   43 &   45.36 &   11 &   31 &    0.5 &    110. &   -105. &      845. &    -1.0 \\
\enddata

Notes. The $x,G$ and $y,G$ coordinates are in arcsec, measured from the center of M 60. Their
orientation is described in the text. A value of $-1$ for $m$(5007) indicates that it was not 
measured.

\end{deluxetable}

\begin{deluxetable}{rccrccrrrrr}
\tablecaption{PN Candidates more likely to belong to NGC 4647\label{tbl-4}}
\tablewidth{0pt}
\rotate
\tabletypesize{\small}
\tablehead{

\colhead{ID} &
\multicolumn{3}{c}{$\alpha$} & \multicolumn{3}{c}{$\delta$} & 
\colhead{$x$, G} & \colhead{$y$, G} & \colhead{Helioc. RV} & \colhead{$m$} \\

\colhead{} & 
\multicolumn{3}{c}{(J2000)} & \multicolumn{3}{c}{(J2000)} & 
\colhead{} & \colhead{} & \colhead{(km s$^{-1}$)} & \colhead{(5007) }}

\startdata
 1029  &   12 & 43 & 30.14  &   11 & 32 & 37.6 &   -131.6 &  -68.8 &   1419.4 & -1.0   \\
 1050  &   12 & 43 & 32.43  &   11 & 33 & 21.4 &   -110.4 &  -17.8 &   1477.7 & -1.0   \\
    3  &   12 & 43 & 33.09  &   11 & 35 & 51.3 &   -139.8 &  129.5 &   1433.5 & 27.3   \\
    4  &   12 & 43 & 34.03  &   11 & 35 & 40.3 &   -123.7 &  122.5 &   1463.0 & 27.6   \\
    6  &   12 & 43 & 34.73  &   11 & 35 & 42.4 &   -114.3 &  127.2 &   1415.2 & 26.8   \\
    9  &   12 & 43 & 35.16  &   11 & 35 & 56.4 &   -111.8 &  142.3 &   1424.1 & 26.6   \\
   11  &   12 & 43 & 35.28  &   11 & 34 &  9.9 &    -82.5 &   39.9 &   1479.6 & 26.9   \\
   14  &   12 & 43 & 35.88  &   11 & 34 &  8.7 &    -73.7 &   41.0 &   1480.8 & 26.6   \\
   17  &   12 & 43 & 36.22  &   11 & 34 & 29.7 &    -74.3 &   62.6 &   1437.8 & 27.3   \\
   21  &   12 & 43 & 36.46  &   11 & 34 & 58.1 &    -78.3 &   90.9 &   1471.1 & 27.0   \\
   22  &   12 & 43 & 36.79  &   11 & 34 & 58.5 &    -73.6 &   92.6 &   1448.2 & 26.8   \\
   24  &   12 & 43 & 37.01  &   11 & 36 &  0.2 &    -86.5 &  153.0 &   1423.1 & 27.0   \\
   25  &   12 & 43 & 37.20  &   11 & 33 & 52.5 &    -50.7 &   30.4 &   1444.0 & 26.8   \\
   27  &   12 & 43 & 37.61  &   11 & 35 & 18.4 &    -67.2 &  114.9 &   1464.0 & 27.1   \\
   31  &   12 & 43 & 38.06  &   11 & 35 & 42.4 &    -67.0 &  139.8 &   1498.5 & 27.7   \\
   33  &   12 & 43 & 38.16  &   11 & 35 &  1.1 &    -54.9 &  100.3 &   1406.8 & 26.6   \\
   34  &   12 & 43 & 38.38  &   11 & 35 &  1.4 &    -51.8 &  101.4 &   1465.2 & 27.0   \\
   37  &   12 & 43 & 38.66  &   11 & 34 & 39.4 &    -42.2 &   81.2 &   1447.5 & 27.5   \\
   39  &   12 & 43 & 38.76  &   11 & 35 & 48.7 &    -58.7 &  148.6 &   1478.4 & 27.3   \\
   47  &   12 & 43 & 39.24  &   11 & 34 & 32.7 &    -32.2 &   77.0 &   1445.3 & 27.1   \\
   50  &   12 & 43 & 39.36  &   11 & 35 & 43.8 &    -48.9 &  146.1 &   1433.2 & 27.2   \\
   54  &   12 & 43 & 39.46  &   11 & 34 & 36.8 &    -30.2 &   81.8 &   1407.3 & 27.8   \\
   60  &   12 & 43 & 39.70  &   11 & 35 &  9.2 &    -35.1 &  114.0 &   1431.7 & 27.0   \\
   62  &   12 & 43 & 39.74  &   11 & 35 &  2.0 &    -32.7 &  107.2 &   1433.0 & 26.5   \\
   72  &   12 & 43 & 40.29  &   11 & 34 &  5.8 &    -10.3 &   55.0 &   1438.0 & 27.1   \\
   99  &   12 & 43 & 41.62  &   11 & 35 & 22.8 &    -11.4 &  134.4 &   1451.4 & 27.2   \\
  109  &   12 & 43 & 41.98  &   11 & 36 &  1.4 &    -16.3 &  173.1 &   1440.1 & 27.1   \\
  120  &   12 & 43 & 42.48  &   11 & 34 & 46.2 &     10.3 &  102.4 &   1448.7 & 26.4   \\
\enddata

Notes. The $x,G$ and $y,G$ coordinates are in arcsec, measured from the center of M 60. Their
orientation is described in the text. A value of $-1$ for $m$(5007) indicates that it was not 
measured.

\end{deluxetable}

\end{document}